\documentclass[a4paper,11pt]{article}
\usepackage{jheppub}

\usepackage[T1]{fontenc}
\usepackage{amssymb}
\usepackage{amsmath}
\usepackage{graphicx}
\usepackage{subcaption}
\usepackage[colorlinks=true]{hyperref}
\usepackage{appendix}
\usepackage{amsbsy}
\usepackage{slashed}
\usepackage{tikz}
\usepackage{tikz-feynman}
\usepackage{multicol}

\newcommand{\code}[1]{\textsc{#1}}

\DeclareMathOperator{\Tr}{Tr}

\preprint{{\raggedleft
		ZU-TH 07/23, TUM-HEP 1451/23\\
}}

\title{\boldmath Two-loop helicity amplitudes for $H+$jet production to higher orders in the dimensional regulator}

\author[a]{Thomas Gehrmann,}
\author[a]{Petr Jakub\v{c}\'{i}k,}
\author[b]{Cesare Carlo Mella,}
\author[b]{Nikolaos Syrrakos,}
\author[b]{Lorenzo Tancredi}

\affiliation[a]{Physik-Institut, Universit\"{a}t Zurich, 
            Winterthurerstrasse 190,
           	CH-8057 Z\"{u}rich,
            Switzerland}
\affiliation[b]{Physik Department, Technische Universit\"{a}t M\"{u}nchen, James-Franck-Straße 1, 85748 Garching,
Germany}
            
\emailAdd{thomas.gehrmann@physik.uzh.ch}            
\emailAdd{petr.jakubcik@physik.uzh.ch}
\emailAdd{cesarecarlo.mella@tum.de}
\emailAdd{nikolaos.syrrakos@tum.de}
\emailAdd{lorenzo.tancredi@tum.de}

\abstract{
In view of the forthcoming High-Luminosity phase of the LHC, next-to-next-to-next-to-leading (N$^3$LO) calculations for the most phenomenologically relevant processes become necessary. In this work, we take the first step towards this goal for H$+$jet production by computing the one- and two-loop helicity amplitudes for the two contributing processes, $H\to ggg$, $H\to q\bar{q}g$, in an effective theory with infinite top quark mass, to higher orders in the dimensional regulator.  We decompose the amplitude in scalar form factors related to the helicity amplitudes and in a new basis of tensorial structures. The form factors receive contributions from Feynman integrals which were reduced to a novel canonical basis of master integrals. We derive and solve a set of differential equations for these integrals in terms of Multiple Polylogarithms (MPLs) of two variables up to transcendental weight six. }

\begin{document}
       
\maketitle

\section{Introduction}\label{sec:intro}
A little over a decade ago, the Higgs boson was discovered after analysing the data collected during Run I at the Large Hadron Collider (LHC) at CERN~\cite{ATLAS:2012yve,CMS:2012qbp}. 
The discovery was achieved while the collider was running at reduced centre-of-mass energies of 7 and 8~TeV and with only a small fraction of the total dataset which will be accumulated 
during its entire runtime. Indeed, it is expected that the forthcoming High-Luminosity phase 
of the LHC (HL-LHC) will yield a dataset corresponding to 3 $ab^{-1}$ 
of integrated luminosity for $pp$ collisions at 14~TeV~\cite{Dainese:2019rgk}. 

Since its discovery, the Higgs boson has been at the centre of the experimental effort at the LHC~\cite{Heinrich:2020ybq}. Studying its properties improves our understanding of electroweak symmetry breaking (EWSB), 
the mechanism which is believed to be
 responsible for the generation of the masses of fermions and weak gauge bosons. 
The dominant production channel for the Higgs boson at the LHC is gluon fusion. 
In the Standard Model, the Higgs coupling to two gluons is mainly mediated through 
a loop of top quarks, making it a loop-induced process already at Leading Order (LO).
For this reason, 
computing higher order perturbative corrections to Higgs production in the full theory quickly becomes prohibitive.

It was realized long ago that radiative corrections can increase the LO Higgs cross-section
in gluon fusion by as much as $\mathcal{O}(100 \%)$~\cite{Dawson:1990zj,Spira:1995rr} and describing its production in hadron collisions thus requires higher-order calculations. Indeed, quite recently the computation of the fully inclusive Higgs cross-section with full dependence on the top quark mass has 
been pushed to next-to-next-to-leading-order (NNLO)~\cite{Czakon:2021yub} 
using numerical techniques to handle the required two- and three-loop scattering amplitudes.

A useful alternative to performing calculations with full dependence on the top mass
is to work in the heavy top quark mass limit $M_{t}\to \infty$. 
Under the assumption that the top quark is the largest scale involved in the calculation,
one can integrate out the top mass and formulate an effective 
Lagrangian for the $Hgg$ coupling~\cite{Wilczek:1977, Shifman:1978, Inami:1982xt}.
In this description, the top-quark loop mediating the $Hgg$ interaction shrinks 
to a point and calculations start at tree level, involving only massless partons. In this limit, inclusive~\cite{Anastasiou:2015vya, Mistlberger:2018etf, Anastasiou:2016cez} 
as well as fully differential~\cite{Chen:2021isd} predictions for Higgs boson production 
via gluon fusion are known up to 
next-to-next-to-next-to-leading-order (N$^3$LO). 

One of the most promising observables to study the EWSB mechanism is the Higgs transverse momentum, see for example~\cite{Bishara:2016jga}.
To remain differential in external radiation, one must study the production of a Higgs boson in association with (at least) one resolved jet. For this process, results in the heavy top quark limit are 
currently known up to NNLO~\cite{Boughezal:2015dra,Boughezal:2015aha,Caola:2015wna,Chen:2016zka,Campbell:2019gmd},
and the residual theoretical uncertainty can be estimated at around $\mathcal{O}(5\%)$.
The heavy top quark limit approximation is valid for transverse momentum of the Higgs which is lower than two times the top quark mass, $p_T < 2 \, m_t$. For higher $p_T$, the heavy quark loop is resolved and finite mass corrections are needed. Results with finite top mass at NLO were first obtained for the very high transverse momentum kinematic region, $p_T \gg 2 m_t $,~\cite{Lindert:2018iug}, and numerically for general kinematics~\cite{Jones:2018hbb,Chen:2021azt}. More recently, following the calculation of the relevant two-loop master integrals~\cite{Bonciani:2016qxi,Bonciani:2019jyb}, the full NLO analytical calculation has been completed~\cite{Bonciani:2022jmb}. 

Taking into account the wealth of currently available data, as well as the expected high-luminosity phase, N$^3$LO calculations will become essential~\cite{Caola:2022ayt}
in order to perform phenomenological studies at the $1\%$ level at the LHC.
A key ingredient to extend the current calculations to N$^3$LO are  
the four-point amplitudes for the production of a Higgs boson and a parton in parton-parton collisions. Working in the
effective theory described above, 
one needs to compute tree level, one-loop, two-loop and three-loop amplitudes for the
scattering of three massless partons and one massive scalar. 
On top of their phenomenological importance, the structure of these 
amplitudes to higher loops is also of formal interest and has been the subject of thorough investigation.
In this context it is worth noticing that contrary to the naive expectation, up to two loops, the finite remainder amplitudes 
for the decay of a Higgs boson to three gluons
have been shown to be expressible in terms of just classical polylogarithms~\cite{Duhr:2012fh}.

Starting at one loop,  amplitudes exhibit singularities which can be regulated in the framework of dimensional regularization. In $D=4-2\epsilon$ dimensions, 
the loop amplitudes are computed as a Laurent expansion in $\epsilon$. An N$^3$LO computation requires one-loop amplitudes 
up to $\mathcal{O}(\epsilon^4)$, two-loop amplitudes up to $\mathcal{O}(\epsilon^2)$
and three-loop amplitudes up to $\mathcal{O}(\epsilon^0)$. 
Our goal in this paper is to provide the first ingredient for such a calculation, namely the two-loop amplitudes up to order $\epsilon^2$.
These amplitudes 
were previously obtained in~\cite{Gehrmann:2012}
to order $\epsilon^0$, providing a key
ingredient to the calculation of NNLO 
QCD corrections to Higgs+jet production~\cite{Boughezal:2015dra,Boughezal:2015aha,Caola:2015wna,Chen:2016zka,Campbell:2019gmd} 
and 
the Higgs boson transverse momentum
distribution~\cite{Chen:2016zka,Chen:2018pzu}.

While our overall approach is relatively standard, we include multiple new elements which help us organize the calculation more efficiently in view of a subsequent extension to three loops.
First of all, we cast the relevant amplitudes into a compact tensorial basis and construct new helicity projectors to extract the corresponding helicity amplitudes directly from 
the Feynman diagrams~\cite{Peraro:2019cjj, Peraro:2020sfm}. We then employ standard integration-by-parts identities~\cite{Chetyrkin:1981qh,Laporta:2000dsw} 
to express these amplitudes in terms of so-called master integrals, and evaluate them using the differential equation method~\cite{Kotikov:1990kg, Kotikov:1991hm, Kotikov:1991pm, Gehrmann:1999as}.
At two loops, the master integrals were computed up to transcendental weight four more than two decades ago~\cite{Gehrmann:2000zt, Gehrmann:2001ck}. 
 Since then, substantial advances were made in the understanding of mathematical structures underlying Feynman integrals and their associated differential equations.

Indeed, about a decade ago it was realized that a special class of Feynman integrals, dubbed \emph{local integrals}~\cite{Arkani-Hamed:2010pyv,Kotikov:2010gf}, 
plays a crucial role in representing scattering amplitudes,
in particular in the case of (planar) $N=4$ Super Yang-Mills theory (SYM). These integrals only feature singularities of the logarithmic type and exhibit
uniform maximum transcendentality~\cite{Kotikov:2006ts,Kotikov:2010gf}, which has for a long time been conjectured to characterize scattering amplitudes in $N=4$ SYM~\cite{Kotikov:2001sc}.
While these properties do not translate in an obvious way to non-supersymmetric theories like QCD, it has been shown that such integrals can still substantially simplify 
the calculations of scattering amplitudes within the Standard Model. Namely, integrals of this type fulfil particularly simple systems of differential equations, in the so-called 
canonical form~\cite{Henn:2013pwa}. Canonical sets of equations are more easily solved and provide a direct handle on the analytic properties of the corresponding integrals in various singular regions. Importantly, the solution of a canonical system with rational coefficients is straightforwardly expressed in terms
of a well-understood class of functions, the Multiple Polylogarithms (MPLs)~\cite{Goncharov:1998kja,Remiddi:1999ew,Vollinga:2004sn}.

Among the initial applications of this formalism 
were planar ladder-type integrals in $H+$jet production up to three loops~\cite{DiVita:2014pza}, which include the planar
two-loop integrals up to $\mathcal{O}(\epsilon^2)$
as a subset.  Here we consider the full set of two-loop integrals: planar and non-planar. We construct a pure basis of uniform transcendental weight and demonstrate that it can be solved in terms of MPLs to any order in $\epsilon$.
At variance with~\cite{Gehrmann:2000zt, Gehrmann:2001ck}, we show that using a generalized 
set of regularity conditions on the canonical master integrals,
all boundary conditions required to fix the solution of the differential equations can be inferred in terms of a small number of one-scale two- and three-point functions which are known in closed form in the literature.
In view of applications at N$^3$LO, we limit ourselves to perform the calculation explicitly to order $\epsilon^2$, which corresponds to transcendental weight six.

The structure of the paper is as follows. The effective coupling of the Higgs boson to light partons and the definition of kinematics is given in Section~\ref{sec:notation}. In Section~\ref{sec:amps}, we formulate the general structure of the amplitude and use projectors to obtain tensor coefficients and construct the helicity amplitudes. 
In Section \ref{sec:integrals}, we describe in detail the construction of pure bases for the two-loop integral families and solve their canonical differential equations analytically. The ultraviolet renormalization and the subtraction of infrared singularities of our amplitudes are discussed in Section \ref{sec:renorm}. Finally, the crossing of the helicity amplitudes to all appropriate kinematic configurations and the necessary analytic continuation of the relevant multiple polylogarithms is discussed in Section \ref{sec:ancont}. We conclude with a brief summary 
in  Section~\ref{sec:conclusions}.

\section{Notation and kinematics}\label{sec:notation}

\subsection{The effective Lagrangian}
The Higgs boson interacts with Standard Model particles with a coupling strength proportional to their mass, and therefore cannot couple directly to gluons or massless quarks. Nevertheless, starting at one loop, the Higgs can interact with gluons through virtual loops of massive quarks. 
In the limit of a very heavy quark mass, $m_q \rightarrow \infty$, one can show that this coupling becomes independent of $m_q$ and an effective theory can be formulated by integrating out the corresponding quark from the full theory~\cite{Wilczek:1977, Shifman:1978, Inami:1982xt}. 
While most quarks have relatively small masses compared to the typical energy scales of scattering processes at the LHC and can often be considered as massless, the same is not true for the top quark. Since it is the heaviest of all known Standard Model particles, this 
effective field theory (EFT) works extremely well for the top quark, at least as long as all scales involved are smaller than twice its mass~\cite{Ellis:1988,Bauri:1990}.  

In the following, we will work in the EFT with a matter content of $N_f=5$ massless quarks and one very heavy quark, the top quark, integrated out. 
In this case, the effective Lagrangian becomes
\begin{align}
    \mathcal{L}_{int} = -\, \frac{\lambda}{4} H G_a^{\mu \nu} G_{a, \mu \nu},
    \label{eqn: effective lagrangian}
\end{align}
where $G_a^{\mu \nu}$ is the field strength tensor of the gluons and $H$ is the Higgs field. From dimensional analysis, the effective coupling $\lambda$ has inverse mass dimension. It was shown long ago~\cite{Spira:95,Chetyrkin:98} how to perform the matching of this effective theory to the Standard Model Lagrangian~\cite{SpiraHiggs:1993, Spira:1995rr, SpiraHiggs3:1998}.

\subsection{Kinematics}
We are ultimately interested in computing the amplitude for the production of a Higgs boson 
and a hadronic jet in parton-parton annihilation at the LHC. 
For simplicity, we start considering the problem in the crossed kinematics which corresponds to
the decay of a Higgs boson into three partons. There are two relevant partonic channels,
namely the decay into three gluons
\begin{equation}
H(p_4) \rightarrow g_1(p_1) + g_2(p_2) + g_3(p_3),
\end{equation}
and into a quark, anti-quark and a gluon
\begin{equation}
H(p_4) \rightarrow q(p_1) + \bar{q}(p_2) + g(p_3).
\end{equation}
The amplitudes for the production processes can then be obtained through 
an analytic continuation of the decay kinematics~\cite{Gehrmann:2002zr}, see Section \ref{sec:ancont}
for more details.

The Mandelstam invariants are defined as
\begin{equation}
s_{12} = (p_1+p_2)^2\,, \quad \quad  s_{13} = (p_1+p_3)^2\,, \quad \quad s_{23} = (p_2+p_3)^2,
\end{equation}
and satisfy the conservation equation
\begin{equation}
s_{12} + s_{13} + s_{23} = M_H^2,
\label{eqn:mandelstamsum}
\end{equation}
where $M_H$ is the mass of the Higgs particle. It is more convenient to work with dimension-less ratios
\begin{equation}
    x = \frac{s_{12}}{M_H^2}, \quad \quad  y = \frac{s_{13}}{M_H^2}, \quad \quad z = \frac{s_{23}}{M_H^2}, 
    \label{eqn:xyz}
\end{equation}
such that \eqref{eqn:mandelstamsum} implies the relation,
\begin{equation}
x+y+z=1.
\label{eqn:xyzsum}
\end{equation}
In the decay kinematic region, all these invariants are non-negative. This, together with \eqref{eqn:xyz}, defines the corresponding kinematic region
\begin{equation}
 z \geq 0 , \quad \quad  0 \leq  y  \leq 1 - z, \quad  \quad x = 1 - y -z. 
\end{equation}

\section{Tensor Decomposition}\label{sec:amps}
Following earlier work on this process~\cite{Gehrmann:2012}, we decompose the amplitudes $\mathcal{M}_{ggg}$ and $\mathcal{M}_{q\overline{q}g}$ as
\begin{equation}
\begin{aligned}
    \mathcal{M}_{ggg} &= \mathcal{S}_{\mu \nu \rho}(g_1,g_2,g_3) \, \epsilon_1^{\mu} \epsilon_2^{\nu} \epsilon_3^{\rho}, \\
    \mathcal{M}_{q\overline{q} g} &= \mathcal{T}_{\mu}(q, \bar{q}, g ) \, \epsilon^{\mu},
\end{aligned}
\label{eqn: amps1}
\end{equation}
where we used $\epsilon_i$ to denote the polarization vectors of external gluons.
The above tensors can be expanded perturbatively in the QCD coupling constant $\alpha_s$ as
\begin{align}
    \mathcal{S}_{\mu \nu \rho}(g_1,g_2,g_3) =\, &\lambda \sqrt{4 \pi \alpha_s} f^{a_1 a_2 a_3} \Big[ \mathcal{S}^{(0)}_{\mu \nu \rho}(g_1,g_2,g_3) + \Big( \frac{\alpha_s}{2 \pi} \Big) \, \mathcal{S}^{(1)}_{\mu \nu \rho}(g_1,g_2,g_3) \nonumber \\  &\phantom{\lambda \sqrt{4 \pi \alpha_s} f^{a_1 a_2 a_3}\;}+ \Big( \frac{\alpha_s}{2 \pi} \Big)^2 \, \mathcal{S}^{(2)}_{\mu \nu \rho}(g_1,g_2,g_3) + \mathcal{O}(\alpha_s^3) \Big]\,, \label{eqn:tensorggg} \\ 
     \mathcal{T}_{\mu }(q,\overline{q},g) =\, &\lambda \sqrt{4 \pi \alpha_s} T^{a}_{ij} \Big[ \mathcal{T}^{(0)}_{\mu}(q,\overline{q},g) + \Big( \frac{\alpha_s}{2 \pi} \Big) \, \mathcal{T}^{(1)}_{\mu}(q,\overline{q},g) \nonumber \\ &\phantom{\lambda \sqrt{4 \pi \alpha_s} T^{a}_{ij}\;} +\Big( \frac{\alpha_s}{2 \pi} \Big)^2 \, \mathcal{T}^{(2)}_{\mu}(q,\overline{q},g) + \mathcal{O}(\alpha_s^3) \Big]\,, \label{eqn:tensorqqg}
\end{align}
where the coefficients $\mathcal{S}^{(i)}_{\mu \nu \rho}$ and $\mathcal{T}_\mu^{(i)}$ are the $i$-loop  contributions to the amplitude. The $SU(3)$ group generators are normalised as $\Tr(T^a T^b) = \delta^{ab}/2$.

Given the external states and their possible helicity and spin configurations, the amplitudes $\mathcal{S}_{\mu \nu \rho}$ and $\mathcal{T}_{\mu}$ can only depend on a limited number of tensor structures. These structures can be further constrained by exploiting symmetries and choosing a gauge or, following~\cite{Gehrmann:2012}, by enforcing gauge invariance through the Ward identities. In this Section, we aim to find such a tensor basis in order to be able to work with the scalar coefficients of the amplitudes with respect to this basis, known as the form factors. The form factors, in turn, are obtained by applying projector operators on the full amplitude expanded in Feynman diagrams. We derive a basis of 4 tensor structures for $H \rightarrow ggg$ and a basis of 2 tensor structures for $H \rightarrow q\bar{q}g$, and the corresponding projectors in Subsections \ref{subsec: TensorDec ggg} and \ref{subsec: TensorDec qqg}, respectively. 

Since we are ultimately interested in fixing the helicities of external states and computing the associated helicity amplitudes, we introduce the spinor-helicity formalism in Subsection \ref{subsec: HelAmpl} and derive a set of genuinely independent helicity amplitudes. These are written as a unique spinor factor times a scalar coefficient which is a linear combination of the form factors. The same linear combination of form factor projectors also defines a helicity amplitude projector.

\subsection{Tensor decomposition for \(H \rightarrow ggg \)}\label{subsec: TensorDec ggg}
Let us start considering the decay of a scalar boson into three massless spin-one particles. The most general tensor structure one can build using the four-vectors associated with the external particles is
\begin{align}
    \mathcal{S}_{\mu \nu \rho}(g_1,g_2,g_3) \epsilon_1^{\mu}\epsilon_2^{\nu}\epsilon_3^{\rho} &= \, \sum_{i,j,k=1}^{3} A_{ijk}  \, p_i \cdot \epsilon_1 \, p_j \cdot \epsilon_2 \,  p_k \cdot \epsilon_3  + \sum_{i=1}^3 B_i \, p_i \cdot \epsilon_1 \epsilon_2 \cdot \epsilon_3 \nonumber \\& \phantom{ = \;} +\sum_{i=1}^3 C_i \, p_i \cdot \epsilon_2 \epsilon_3 \cdot \epsilon_1 +\sum_{i=1}^3 D_i \, p_i \cdot \epsilon_3 \epsilon_1 \cdot \epsilon_2 \,.\nonumber 
    \end{align}
    There are four helicity configurations for this amplitude, so we expect four independent tensor structures in $D=4$. Indeed, the transversality conditions $p_i \cdot \epsilon_i = 0$, $i=1,2,3$ and the cyclic gauge choice $\epsilon_1 \cdot p_{2} = 0$, $\epsilon_2 \cdot p_{3} = 0$, $\epsilon_3 \cdot p_{1} = 0$ restrict the tensor structures considerably:
    \begin{align}
   \mathcal{S}_{\mu \nu \rho}(g_1,g_2,g_3) \epsilon_1^{\mu}\epsilon_2^{\nu}\epsilon_3^{\rho} &= A_{312} \, p_3  \cdot \epsilon_1 \, p_1 \cdot \epsilon_2 \, p_2 \cdot \epsilon_3 + B_3 \, \epsilon_2 \cdot \epsilon_3 \, p_3 \cdot \epsilon_1 \nonumber \\ & \phantom{=\,\,} + C_1 \, p_1 \cdot \epsilon_2 \, \epsilon_3 \cdot \epsilon_1 + D_2 \, p_2 \cdot \epsilon_3 \, \epsilon_1 \cdot \epsilon_2\,,\label{eqn:gggTensorDec0}\\
    &=  \mathcal{G}_1 T_1 + \mathcal{G}_2 T_2 + \mathcal{G}_3 T_3 + \mathcal{G}_4 T_4\,, 
 \label{eqn:gggTensorDec1}
\end{align}
where we relabelled the coefficients $\{A_i, ..., D_i\}$ as the form factors $\mathcal{G}_i$ and defined the basis
\begin{align}
T_1 &=  p_1 \cdot \epsilon_2 \,\,  \epsilon_3 \cdot \epsilon_1,
\nonumber \\
T_2 &=  p_2 \cdot \epsilon_3 \, \,\epsilon_1 \cdot \epsilon_2, \nonumber \\
T_3 &= \epsilon_2 \cdot \epsilon_3 \,\,  p_3 \cdot \epsilon_1, \nonumber \\
T_4 &= p_3  \cdot \epsilon_1 \, p_1 \cdot \epsilon_2 \, p_2 \cdot \epsilon_3 \,.
\label{eqn:gggTensors}
\end{align}
The form factors can be obtained from a Feynman diagram decomposition of the amplitude $\mathcal{S}^{(i)}_{\mu \nu \rho}$ at any loop order by applying suitable projectors $\mathcal{P}_i$ defined as
\begin{align}
\sum_{pol} \mathcal{P}_i \, \mathcal{S}_{\mu \nu \rho}(g_1,g_2,g_3) \epsilon_1^{\mu}\epsilon_2^{\nu}\epsilon_3^{\rho} = \mathcal{G}_i\,,
\label{eqn:gggSumPol}
\end{align}
where polarization vectors satisfy the gauge-fixed polarization sum, with reference vectors defined above. 
The projectors can in turn be decomposed in terms of the dual of the tensor basis in \eqref{eqn:gggTensorDec1}:
\begin{align} \label{eqn:projdecomp}
\mathcal{P}_i = \sum_{j=1}^4 c_i^{(j)} T_j^{\dagger}\,.
\end{align}
To work out the projectors explicitly, we
insert the decompositions \eqref{eqn:projdecomp} and \eqref{eqn:gggTensorDec1} into the definition \eqref{eqn:gggSumPol} and obtain the requirement
\begin{align}
\sum_{j=1}^4 c_i^{(j)} T_j^{\dagger}\sum_{k=1}^4 \mathcal{G}_k T_k \overset{!}{=} \mathcal{G}_i\,,
\end{align}
which is satisfied by the coefficients $c_{i}^{(j)} = (T^{\dagger}_jT_i)^{-1}$. In particular for the amplitude $H \rightarrow ggg$ with external states in $D$~dimensions~\cite{Peraro:2020sfm}, we get
\begin{align}
    \mathcal{P}_1 &= \frac{1}{s_{12} s_{13} (D-3)} (s_{23} \, T_1^{\dagger} - T_4^{\dagger}), \nonumber \\ 
    \mathcal{P}_2 &= \frac{1}{s_{12}s_{23}(D-3)} (s_{13} \, T_2^{\dagger} - T_4^{\dagger}), \nonumber \\ 
    \mathcal{P}_3 &= \frac{1}{s_{13}s_{23}(D-3)} (s_{12} \, T_3^{\dagger} - T_4^{\dagger}), \nonumber \\ 
    \mathcal{P}_4 &= \frac{1}{s_{12}s_{13}s_{23}(D-3)} (D \, T_4^{\dagger} - s_{12} T_3^{\dagger} - s_{23} T_1^{\dagger} - s_{13} T_2^{\dagger})\,.
    \label{eqn:gggFproj}
\end{align}
\subsection{Tensor decomposition for $H \rightarrow q\bar{q}g$}\label{subsec: TensorDec qqg}
Similarly, it is easy to see that the most general tensor decomposition for two external spinors and a four-vector is
\begin{align}
 \mathcal{T}_{\mu }(q,\overline{q},g) \epsilon_3^{\mu} &=
\sum_{\Gamma} A_{\Gamma} \, \,  \bar{u}(p_1)\, \Gamma^{}_{\mu} \, v(p_2) \epsilon_3^{\mu} \nonumber \\
&+ \sum_{j=1,2,3} \, \sum_{\Gamma'} B_{\Gamma'} \,  \bar{u}(p_1)\, \Gamma' \, v(p_2) \epsilon_3 \cdot p_j
\,,\label{eqn:qqbgTensorDec0}
\end{align}
where we sum over odd products of gamma matrices. In the first sum, all indices but one are contracted amongst each other or with external momenta while in the second, there are no indices left, and the polarization vector contracts with an external momentum vector.\\
In general, one might expect the length of the spinor chains to be bound by the loop order. By simple enumeration, one can see that at tree level there can only be one gamma matrix, 
at one loop up to three gamma matrices and at two loops up to five. Nonetheless, it is easy to see that with the momenta at hand, one cannot build any spinor chain with more than one Dirac $\gamma$ matrix. Enforcing the transversality condition $p_3 \cdot \epsilon_3 = 0$ and gauge choice $\epsilon_3 \cdot p_{1} = 0$, one is  left with the decomposition
\begin{align}
     \mathcal{T}_{\mu }(q,\overline{q},g) \epsilon_3^{\mu} & =\,  \mathcal{F}_1\,  \bar{u}(p_1) \slashed{\epsilon_3} v(p_2) + \mathcal{F}_2 \, \bar{u}(p_1) \slashed{p_3} v(p_2) \epsilon_3 \cdot p_2.
    \label{eqn:qqbgTensorDec1}
\end{align}
Hence we define the two tensor structures
\begin{align}
T_1 &=  \bar{u}(p_1) \slashed{\epsilon_3} v(p_2),\nonumber \\ 
T_2 &=  \bar{u}(p_1) \slashed{p_3} v(p_2)\epsilon_3 \cdot p_2. 
\label{eqn:qqbgTensors}
\end{align}
The form factors are extracted from the amplitude using the projectors $\mathcal{P}_i$ with
\begin{align}
\sum_{pol} \mathcal{P}_i \, \mathcal{T}_{\mu }(q,\bar{q},g) \epsilon_3^{\mu} = \mathcal{F}_i,
\label{eqn:qqbgSumPol}
\end{align}
where polarization vector of the gluon satisfies the gauge-fixed polarization sum, with reference vector defined above.
Following the strategy outlined in Subsection \ref{subsec: TensorDec ggg}, we get for the two projectors
\begin{align}
    \mathcal{P}_1 &=\frac{1}{2s_{12}(D-3)}\Big( T_1^{\dagger} - \frac{1}{s_{23}}T_2^{\dagger}\Big), \nonumber \\ 
    \mathcal{P}_2 &= \frac{1}{2s_{12}s_{23}(D-3)}\Big(\frac{D-2}{s_{23}} T_2^{\dagger} - T_1^{\dagger}\Big).
    \label{eqn:qqbgFproj}
\end{align}
\subsection{Helicity amplitudes}\label{subsec: HelAmpl}
From the tensors found above, we can easily obtain compact expressions for the 
relevant helicity amplitudes. We work in the ’t Hooft-Veltman scheme and consider external states as four-dimensional and assume fixed helicity states. In massless QCD, both gluons and quarks have two helicity configurations, $\lambda_i = \pm$, and the amplitudes can be written as  
\begin{equation}
\begin{aligned}
    \mathcal{M}^{\lambda_1 \lambda_2 \lambda_3 }_{ggg} &= \mathcal{S}_{\mu \nu \rho}(g_1,g_2,g_3) \, \epsilon_{1,\lambda_1}^{\mu}(p_1) \epsilon_{2,\lambda_1}^{\nu}(p_2) \epsilon_{3,\lambda_3}^{\rho}(p_3), \\
    \mathcal{M}_{q\overline{q} g}^{\lambda_1 \lambda_2 \lambda_3 } &= \mathcal{T}_{\mu}(q_{\lambda_1}, \bar{q}_{\lambda_2}, g ) \, \epsilon^{\mu}_{3,\lambda_3}(p_3)\,.
\end{aligned}
\label{eqn: amps2}
\end{equation}
\newline
The two helicity states of a four-component massless spinor are projected out through
\begin{equation}
    \psi_{L} = \frac{1}{2}\big( 1 - \gamma_5 \big) \, \psi, \quad \psi_{R} = \frac{1}{2}\big( 1 + \gamma_5 \big) \, \psi,
\end{equation}
and we fix the spinor-helicity bracket representation of incoming fermions as
\begin{align}
    |p] =\psi_L(p)\,, \quad |p \rangle = \psi_R(p) \,,
    \label{eqn:spinor}
\end{align}
and of incoming anti-fermions as
\begin{align}
    \langle p | = \bar{\psi}_L(p),  \quad [p | = \bar{\psi}_R(p).
    \label{eqn:spinorBar}
\end{align}
For massless vector bosons, incoming states of positive and negative helicity are given by
\begin{equation}
    \epsilon_+^{\mu}(p;r) = -\frac{[ r \, \gamma^{\mu} \, p \rangle}{\sqrt{2}\, [r p]}\,, \qquad  \epsilon_-^{\mu}(p;r) = \frac{\langle r \, \gamma^{\mu} \, p ]}{\sqrt{2}\, \langle r p \rangle},
    \label{eqn:polVec}
\end{equation}
where the reference momentum $r$ is an arbitrary light-like vector such that $r \cdot p \neq 0$. Outgoing states have the same representation with switched helicities.

Let us start by considering the decay $H \rightarrow ggg$. 
There are two independent helicity configurations, which we choose to be $\mathcal{M}^{+++}_{ggg}$ and $\mathcal{M}^{++-}_{ggg}$, while the other helicity amplitudes are obtained through parity conjugation and by relabelling of the gluon momenta.
Starting from the decomposition \eqref{eqn:gggTensorDec1} with the basis \eqref{eqn:gggTensors} and applying the definitions \eqref{eqn:spinor}--\eqref{eqn:polVec}, we can cast the independent helicity amplitudes in terms of spinor products,
\begin{align}
    \mathcal{M}^{+++}_{ggg} &= \alpha \frac{1}{\sqrt{2}}\frac{M_H^4}{\langle 1 \, 2 \rangle \langle 2 \, 3 \rangle \langle 3 \, 1 \rangle}, \\
    \mathcal{M}^{++-}_{ggg} &= \beta \frac{1}{\sqrt{2}}\frac{[1 \, 2]^3}{[2 \, 3 ] [1 \, 3 ]}
    \label{eqn:HA_ggg}\,,
\end{align}
where the coefficients $\alpha$ and $\beta$ are simple linear combinations of 
the original form factors:
\begin{align}
\alpha &=  - \frac{2 s_{12} s_{13} \,\mathcal{G}_1 +2 s_{12} s_{23}\, \mathcal{G}_2 + 2 s_{13} s_{23} \, \mathcal{G}_3 + s_{12} s_{13} s_{23} \,\mathcal{G}_4   }{2 M_H^4}, \\
\beta &= - \frac{2 s_{23} \, \mathcal{G}_2 + s_{23} s_{13} \mathcal{G}_4}{2 s_{12}}.
\label{eqn:alpha beta}
\end{align}
The helicity projectors for $\alpha$ and $\beta$ are built by replacing the $\mathcal{G}_i$  with the corresponding form factor projectors from \eqref{eqn:gggFproj}: 
\begin{align}
\mathcal{P}_{\alpha} &=  - \frac{1}{2 M_H^4} \Big( 2 s_{12} s_{13} \mathcal{P}_1 +2 s_{12} s_{23}\, \mathcal{P}_2 + 2 s_{13} s_{23} \, \mathcal{P}_3 + s_{12} s_{13} s_{23} \,\mathcal{P}_4  \Big)  \nonumber \\
 &= \frac{1}{2 (D-3) M_H^4} \Big( (6 - D) T_4^{\dagger} - s_{23} T_1^{\dagger} - s_{12} T_3^{\dagger} - s_{13} T_2^{\dagger}\Big), \\
\mathcal{P}_{\beta} &=  - \frac{2 s_{23} \, \mathcal{P}_2 + s_{23} s_{13} \mathcal{P}_4}{2 s_{12}} \nonumber \\
&= \frac{1}{2 (D - 3)s_{12}^2} \Big(  s_{23} T_1^{\dagger}  -  s_{13} T_2^{\dagger}  + s_{12} T_3^{\dagger} +  (2 - D) T_4^{\dagger} \Big).
\label{eqn:Projectorsalpha beta}
\end{align}

Let us now consider the other partonic process $H \rightarrow q \bar{q} g$.   
In this case, there is only one independent helicity amplitude, which we choose to be $\mathcal{M}_{q\bar{q}g}^{LR+}$. In this helicity configuration, the first tensor in \eqref{eqn:qqbgTensorDec1} vanishes, while the second tensor yields
\begin{equation}
\mathcal{M}_{q \bar{q} g}^{LR+} = \gamma \frac{1}{\sqrt{2}}\frac{[2 \, 3]^2}{[1 \, 2]},
\label{eqn:HA_qqbg}
\end{equation}
where
\begin{equation}
\gamma = s_{12} \, \mathcal{F}_2.
\label{eqn:gamma}
\end{equation}
Again, the helicity projector is easily obtained by replacing $\mathcal{F}_2$ with the associated projector from \eqref{eqn:qqbgFproj},
\begin{equation}
\mathcal{P}_{\gamma} = s_{12} \, \mathcal{P}_2 =  \frac{1}{2s_{23}(D-3)}\Big(\frac{D-2}{s_{23}} T_2^{\dagger} - T_1^{\dagger}\Big).
\label{eqn:Projectorgamma}
\end{equation}
Just like full amplitudes in (\ref{eqn:tensorggg}) and (\ref{eqn:tensorqqg}), the helicity amplitude coefficients $\alpha$, $\beta$, $\gamma$ also have a perturbative expansion,
\begin{equation}
\Omega = \lambda \sqrt{4 \pi \alpha_s} \,  T_{\Omega} \,  \Big[\Omega^{(0)} + \Big(\frac{\alpha_s}{2 \pi}\Big) \, \Omega^{(1)} + \Big(\frac{\alpha_s}{2 \pi}\Big)^2 \, \Omega^{(2)} + \mathcal{O}(\alpha_s^3) \Big]
\label{eqn:AmplExpansion}
\end{equation}
for $\Omega = \alpha,\beta,\gamma$. The overall colour factors for the two processes are $T_{\alpha}=T_{\beta} = f^{a_1a_2a_3}$ and $T_{\gamma} =T^{a_3}_{\,\,\,i,j}$.
The tree-level helicity amplitudes are well-known and the coefficients evaluate to
\begin{align}
\alpha^{(0)} &= \beta^{(0)} = -1  \\
\gamma^{(0)} &= 1.
\end{align}

\section{Master integrals}\label{sec:integrals}
\begin{table}[b]
  \centering
  \begin{tabular}{c|c}
    \multicolumn{1}{c}{Family PL: $\{P_i\}$} & \multicolumn{1}{c}{Family NPL: $\{N_i\}$} \\
    \hline
    \hline
    $k_1$      &   $k_1$       \\
    $k_2$      &   $k_2$       \\
    $k_1 - k_2$   &  $k_1 - p_1$       \\
    $k_1 - p_1$  &    $k_2-p_1$        \\ 
    $k_2 - p_1$  &    $k_1+p_3$       \\
    $k_1-p_1-p_2$ & $k_2+p_3$ \\
    $k_2-p_1-p_2$ & $k_1+p_2+p_3$\\
    $k_1-p_1-p_2-p_3$ & $k_1-k_2$ \\
    $k_2-p_1-p_2-p_3$ & $k_1-k_2+p_2$ \\
  \end{tabular}
  \caption{The two auxiliary topologies of momenta which label the propagators of every diagram appearing in the amplitudes. PL labels exclusively planar diagrams, whereas NPL accommodates also all non-planar sectors.}\label{tab:auxtopo}
\end{table}
The form factors $\mathcal{F}_i$ and $\mathcal{G}_i$, which relate to the helicity amplitudes via \eqref{eqn:alpha beta}, \eqref{eqn:gamma}, receive contributions from all relevant Feynman diagrams at a given perturbative order. All tree-level, one- and two-loop diagrams with $H$ and $ggg$ or $q\bar{q}g$ external states are generated with standard QCD vertices and the effective Higgs interactions  using \code{QGRAF}~\cite{Nogueira:1991ex}. We shift each diagram to a kinematic crossing of one of the auxiliary topologies presented in Table \ref{tab:auxtopo} using \code{Reduze2}~\cite{Studerus:2009ye,vonManteuffel:2012np}. After inserting Feynman rules and evaluating the Dirac and Lorentz algebra in \code{FORM}~\cite{Vermaseren:2000nd}, the contribution of each diagram to the form factors in \eqref{eqn:gggTensorDec1} and \eqref{eqn:qqbgTensorDec1} can be written as a combination of scalar integrals of the form
\begin{equation}\label{eq:intdef}
    I_{a_{1}, ..., a_{9}} = \int \bigg(\prod_{l=1}^2 (-M_H^2)^{-\epsilon} e^{\gamma_E \epsilon} \frac{\mathrm{d}^D k_l}{i\pi^{d/2}}\bigg) \prod_{i=1}^{9} D_i^{-a_i}\,,
\end{equation}
with $k_l$ the loop momenta, $D_i \in \{P_i\}\cup\{N_i\}$ the internal propagators, and $\gamma_E = 0.577\dots$ the Euler-Mascheroni constant. 
The topology of each scalar integral is uniquely identified by the propagators which appear in the denominator ($a_i\geq 1$). This information can be used to 
compute the diagram's sector ID within one of the auxiliary topologies in
Table~\ref{tab:auxtopo}
as a binary nine-bit number. 
The scalar integrals defined in \eqref{eq:intdef} satisfy integration-by-parts (IBP) identities~\cite{Chetyrkin:1981qh,Laporta:2000dsw}, allowing them to be expressed in terms of a minimal set of so-called master integrals. Note that the physical measure in the bare amplitude is $\mathrm{d}^D k/(2\pi)^{D}$ per loop compared to the integration measure in \eqref{eq:intdef}, which then requires a simple conversion factor when inserting the solutions of the master integrals into the amplitude.

\begin{figure}[t]
    \centering 
\begin{subfigure}{0.28\textwidth}
  \includegraphics[width=\linewidth]{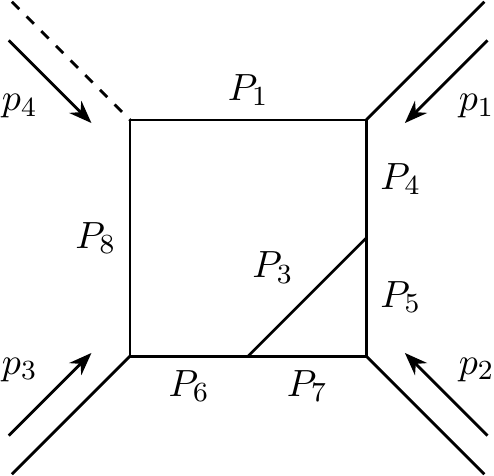}
  \caption{PL 253}
\end{subfigure}\hfil 
\begin{subfigure}{0.28\textwidth}
  \includegraphics[width=\linewidth]{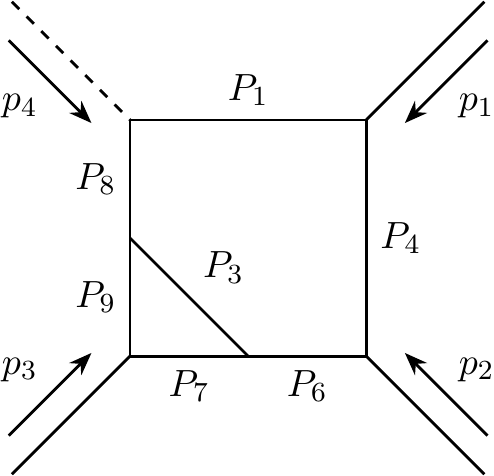}
  \caption{PL 493}
\end{subfigure}\hfil 
\begin{subfigure}{0.28\textwidth}
  \includegraphics[width=\linewidth]{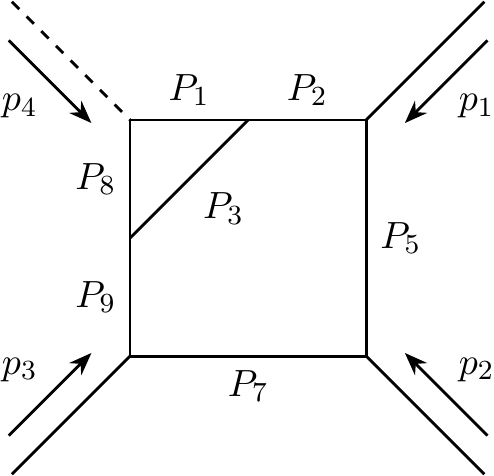}
  \caption{PL 471}
\end{subfigure}\hfil 

\medskip
\begin{subfigure}{0.28\textwidth}
  \includegraphics[width=\linewidth]{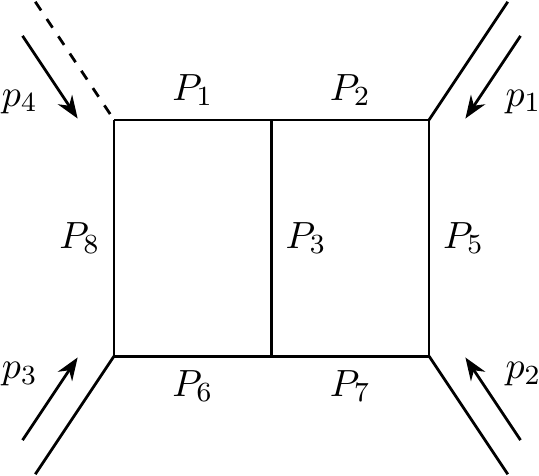}
  \caption{PL 247}
\end{subfigure}\hfil 
\begin{subfigure}{0.28\textwidth}
  \includegraphics[width=\linewidth]{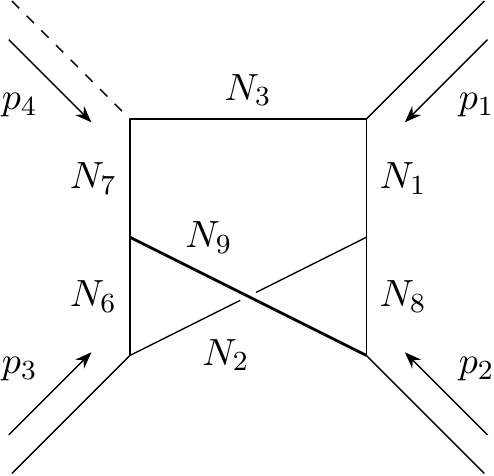}
    \caption{NPL 487}
\end{subfigure}\hfil 
\begin{subfigure}{0.28\textwidth}
  \includegraphics[width=\linewidth]{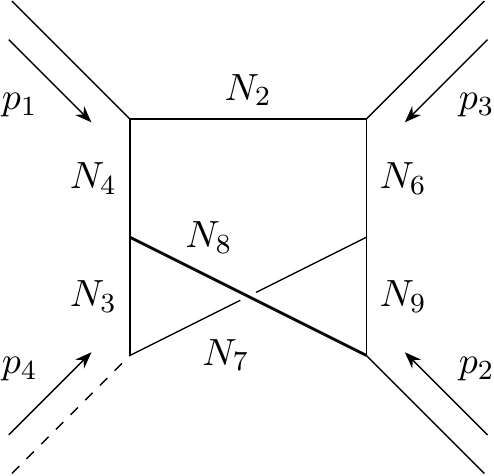}
    \caption{NPLx23 494}
\end{subfigure}\hfil 

\caption{The 4 planar and 2 non-planar sectors with $t=7$, labelled with the propagators $P_{i}$ and $N_{i}$, respectively, as listed in Table \ref{tab:auxtopo}. Their names indicate the integral family and sector ID. The sectors (d)-(f) contain new master integrals and are the focus of this Section. Sectors (a) and (b) are entirely reducible to master integrals found in the trees of the former three top sectors. Sector (c) can be reduced to masters from (d)-(f) and four additional integrals which are easily computed from one-loop results.}
\label{fig:topsectors}
\end{figure}

To evaluate the master integrals, we use the method of differential equations~\cite{Kotikov:1990kg, Kotikov:1991hm, Kotikov:1991pm, Gehrmann:1999as}, augmented with a canonical basis~\cite{Henn:2013pwa} for the master integrals.
In particular, we reduce all scalar integrals appearing in the amplitudes using \code{Reduze2}~\cite{vonManteuffel:2012np} and \code{Kira}~\cite{Maierhofer:2017gsa, Klappert:2020nbg} directly to a canonical basis which involves 89 master integrals. Our basis is different from the one considered in~\cite{Gehrmann:2000zt, Gehrmann:2001ck}, but relations amongst the two sets of integrals can easily be established. There are 4 planar and 2 non-planar sectors with the maximum number of propagators, $t=7$, depicted in Figure \ref{fig:topsectors}. 85 of the 89 IBP master integrals are subsectors of the non-trivial top sectors (d)-(f). They are kinematic crossings of the 16 planar and 8 non-planar topologies depicted in Figures \ref{fig:PLdiags} and \ref{fig:NPLdiags}, which will be explained 
in more detail after introducing the canonical basis. 

\subsection{Canonical basis}\label{subsec:canon}
\begin{figure}[t]
    \centering 
\begin{subfigure}{0.19\textwidth}
  \includegraphics[width=\linewidth]{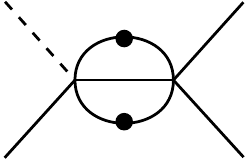}
\end{subfigure}\hfil 
\begin{subfigure}{0.19\textwidth}
  \includegraphics[width=\linewidth]{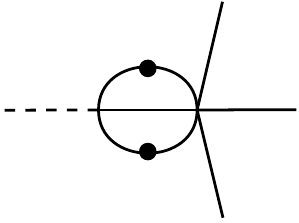}
\end{subfigure}\hfil 
\begin{subfigure}{0.19\textwidth}
  \includegraphics[width=\linewidth]{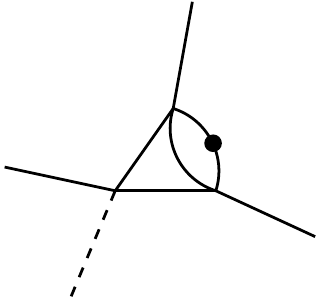}
\end{subfigure}\hfil 
\begin{subfigure}{0.19\textwidth}
  \includegraphics[width=\linewidth]{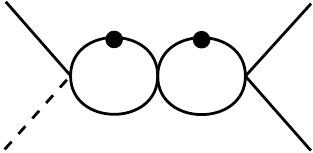}
\end{subfigure}\hfil

\medskip
\begin{subfigure}{0.19\textwidth}
  \includegraphics[width=\linewidth]{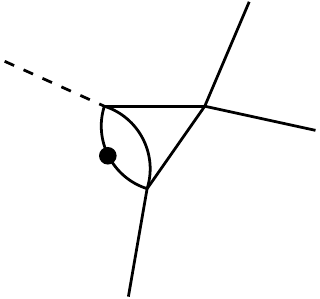}
\end{subfigure}\hfil 
\begin{subfigure}{0.19\textwidth}
  \includegraphics[width=\linewidth]{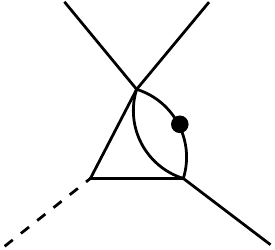}
\end{subfigure}\hfil 
\begin{subfigure}{0.19\textwidth}
  \includegraphics[width=\linewidth]{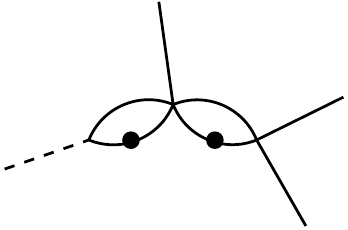}
\end{subfigure}\hfil 
\begin{subfigure}{0.19\textwidth}
  \includegraphics[width=\linewidth]{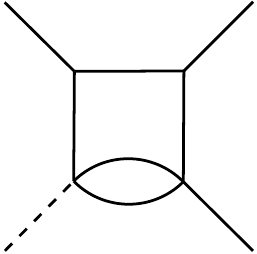}
\end{subfigure}\hfil 

\medskip
\begin{subfigure}{0.19\textwidth}
  \includegraphics[width=\linewidth]{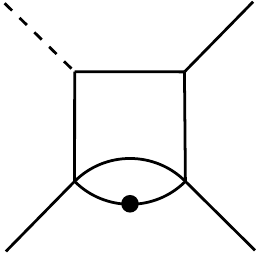}
\end{subfigure}\hfil
\begin{subfigure}{0.19\textwidth}
  \includegraphics[width=\linewidth]{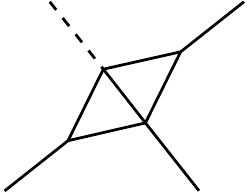}
\end{subfigure}\hfil 
\begin{subfigure}{0.19\textwidth}
  \includegraphics[width=\linewidth]{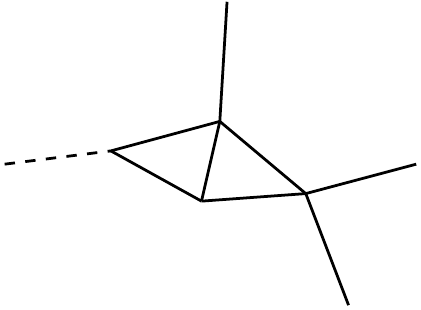}
\end{subfigure}\hfil 
\begin{subfigure}{0.19\textwidth}
  \includegraphics[width=\linewidth]{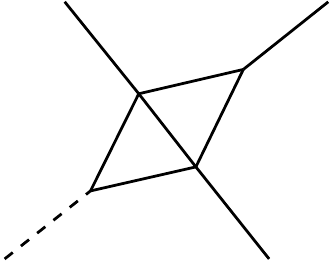}
\end{subfigure}\hfil 

\medskip
\begin{subfigure}{0.19\textwidth}
  \includegraphics[width=\linewidth]{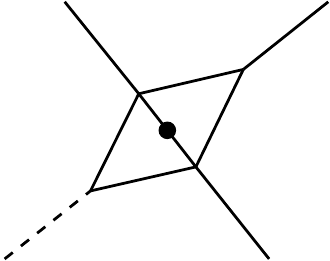}
\end{subfigure}\hfil 
\begin{subfigure}{0.19\textwidth}
  \includegraphics[width=\linewidth]{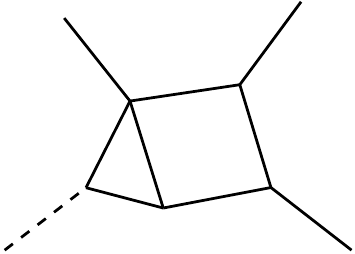}
\end{subfigure}\hfil
\begin{subfigure}{0.19\textwidth}
  \includegraphics[width=\linewidth]{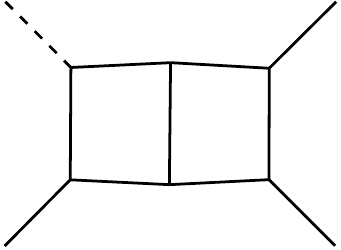}
\end{subfigure}\hfil
\begin{subfigure}{0.19\textwidth}
  \includegraphics[width=\linewidth]{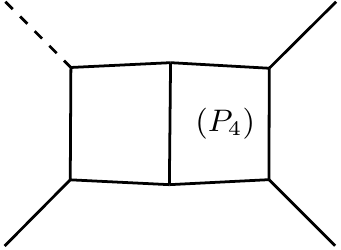}
\end{subfigure}\hfil
\caption{Planar topologies which appear in the canonical basis. The dashed line is the massive leg, dots represent squared propagators. A propagator in brackets denotes an integral with this propagator in the numerator.}
\label{fig:PLdiags}
\end{figure}

\begin{figure}[t]
    \centering 
\begin{subfigure}{0.19\textwidth}
  \includegraphics[width=\linewidth]{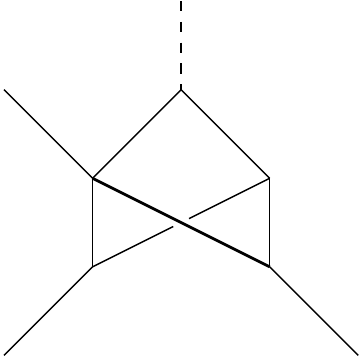}
\end{subfigure}\hfil 
\begin{subfigure}{0.19\textwidth}
  \includegraphics[width=\linewidth]{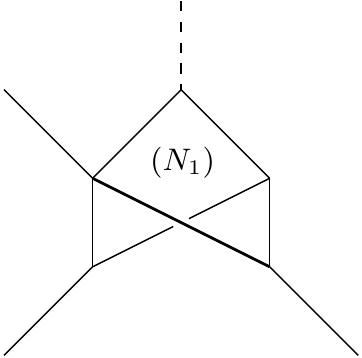}
\end{subfigure}\hfil 
\begin{subfigure}{0.19\textwidth}
  \includegraphics[width=\linewidth]{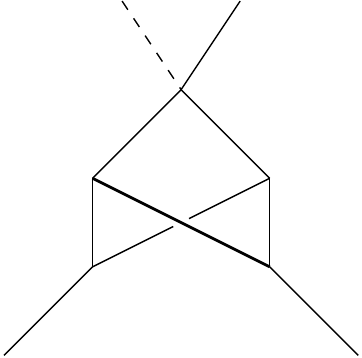}
\end{subfigure}\hfil 
\begin{subfigure}{0.19\textwidth}
  \includegraphics[width=\linewidth]{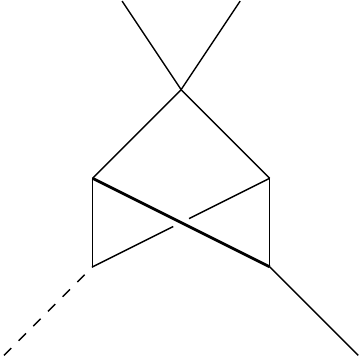}
\end{subfigure}\hfil 
\medskip

\begin{subfigure}{0.19\textwidth}
  \includegraphics[width=\linewidth]{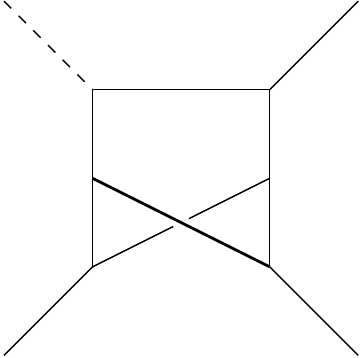}
\end{subfigure}\hfil 
\begin{subfigure}{0.19\textwidth}
  \includegraphics[width=\linewidth]{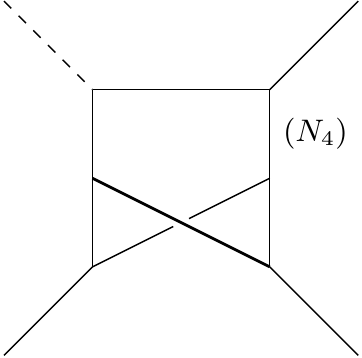}
\end{subfigure}\hfil 
\begin{subfigure}{0.19\textwidth}
  \includegraphics[width=\linewidth]{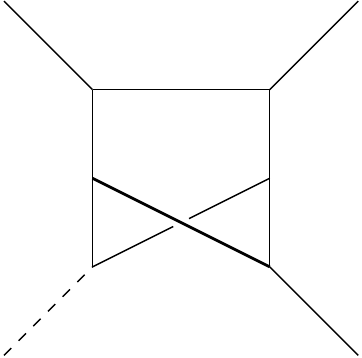}
\end{subfigure}\hfil 
\begin{subfigure}{0.19\textwidth}
  \includegraphics[width=\linewidth]{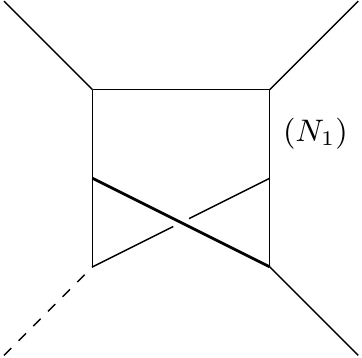}
\end{subfigure}\hfil 

\caption{The non-planar topologies in the canonical basis.}
\label{fig:NPLdiags}
\end{figure}

Multiple public packages exist for the derivation of a candidate canonical basis like \code{CANONICA}~\cite{Meyer:2017joq}, \code{Fuchsia}~\cite{Gituliar:2017vzm}, or \code{DlogBasis}~\cite{Henn:2020lye}.
We use \code{DLogBasis} to provide a list of candidate UV-finite integrals with unit leading singularities for the planar family. For the non-planar family, we constructed canonical candidates by studying the leading singularities of the relevant master integrals, following a loop-by-loop approach. Note that for all planar and some non-planar integrals, canonical candidates can be obtained just by rescaling a single integral in the sector by its maximal cut, i.e.\ a solution to the homogeneous part of its differential equation~\cite{Primo:2016ebd}. This can be confirmed by studying the maximal cut of the specific canonical integral in the Baikov representation~\cite{Baikov:1996iu}. The canonical bases sufficient for the computation of all integrals in the families PL and NPL are presented in the supplementary material. The amplitude, however, contains diagrams which produce integrals in kinematic crossings of the families in Table \ref{tab:auxtopo}. To obtain a basis that is sufficient to represent the physical amplitude, we proceed from the lowest sectors in applying crossings to the canonical integrals in PL and NPL, and appending to our basis those 
canonical integrals whose reduction contains new masters. This yields a minimal set of crossed and uncrossed canonical master integrals, sufficient for physical applications, 
whose generic topologies are depicted in 
Figures~\ref{fig:PLdiags} and~\ref{fig:NPLdiags}. Note that we obtain results first in the Euclidean region, where all invariants are negative $s_{ij} < 0$. For simplicity, we also set $M_{H}^2 = -1$ in explicit formulas below.

Up to now, we only considered the non-trivial top sectors (d)-(f) in Figure \ref{fig:topsectors}, but the amplitude requires the reduction of integrals from all six topologies. While integrals in (a) and (b) can be reduced to integrals considered earlier, (c) contains four new master integrals, $I_{86}, \dots, I_{89}$ which need to be added to the basis. They are the integrals (the numbering refers to the canonical basis in the supplementary material):
\begin{align}
    I_{86} &= \epsilon^2 \left[
\begin{tikzpicture}[baseline=(a1.base), scale = 2]
\begin{feynman}[inline=(a1.base)]
\vertex (x1) at (-1,0.25) {\(p_{1}\)};
\vertex (a1) at (-1,0) {\(p_{2}\)};
\vertex (y1) at (-1,-0.25) {\(p_{3}\)};
\vertex (b1) at (-0.5,0) ;
\vertex (b2) at (0,0);
\vertex (b3) at (0.5,0) ;
\vertex (a3) at (1,0) {\(q\)} ;
\diagram* {
(a1) -- [] (b1),
(x1) -- [] (b1),
(y1) -- [] (b1),
(b1) -- [ half left] (b2),
(b2) -- [ half left] (b1),
(b2) -- [ half left] (b3),
(b3) -- [ half left] (b2),
(b3) -- [scalar] (a3),
};
\vertex [dot] (i) at ($(b1)!0.5!(b2) + (0, 0.21)$) {};
\vertex [dot] (j) at ($(b2)!0.5!(b3) + (0, 0.21)$) {};
\end{feynman}
\end{tikzpicture}\right],\\[3mm] 
    I_{87} &= -\epsilon^3 (1-y-z) z 
     \left[
\begin{tikzpicture}[baseline=(a1.base), scale = 1]
\begin{feynman}[inline=(a1.base)]
\vertex (b1) at (-1,1) {\(p_{3}\)};
\vertex (b2) at (-1,-1) {\(p_{2}\)};
\vertex (b3) at (1,1) {\(q\)};
\vertex (b4) at (1,-1)  {\(p_{1}\)};
\vertex (x1) at (-0.35,0.35) ;
\vertex (x2) at (-0.35,-0.35);
\vertex (x3) at (0.35,0.35) ;
\vertex (x4) at (0.35,-0.35)  ;
\vertex (y3) at (0.55,0.55)  ;
\diagram* {
(b1) -- [] (x1),
(b2) -- [] (x2),
(y3) -- [scalar] (b3),
(x3) -- [half right] (y3),
(x3) -- [half left] (y3),
(b4) -- [] (x4),
(x3) -- [] (x1),
(x1) -- [] (x2),
(x2) -- [] (x4),
(x4) -- [] (x3)
};
\vertex [dot] (i) at ($(x3)!0.5!(y3) + (-0.09, 0.09)$) {};
\end{feynman}
\end{tikzpicture}\right]
\end{align}
and
$$   I_{88} = I_{87}(p_1\leftrightarrow p_2)\,, \qquad
   I_{89} = I_{87}(p_2\leftrightarrow p_3)\,.$$

The full expressions up to $\mathcal{O}(\epsilon^6)$ for $I_{86}$ as well as $I_{87}$ and its two crossings are given in the supplementary material. In this way, one can complete the set of 89 master integrals sufficient to reduce any integral in this process. The full canonical basis is given in the supplementary material accompanying this paper.

\subsection{Solution of differential equations}\label{subsec:DEs}
For the purpose of computing the master integrals, we consider first 
the 16 integrals in the tree of top sector (d) in PL and the 36 integrals in the tree of the non-planar top sectors (e), (f) in NPL. Separately for the two auxiliary topologies, we compute the derivatives of the candidate canonical combinations, and insert the IBP reduction to obtain differential equations in the following form
\begin{align}
    \frac{\partial}{\partial y} \vec{I}(y,z;\epsilon)= \epsilon\left(\frac{1}{y}A_{0}+\frac{1}{y-1}A_{1} + \frac{1}{y+z}A_{z}+\frac{1}{y-(1-z)}A_{1-z} \right)\vec{I}(y,z;\epsilon)\,,\label{eqn: DEy}\\
    \frac{\partial}{\partial z} \vec{I}(y,z;\epsilon)= \epsilon\left(\frac{1}{z}B_{0}+\frac{1}{z-1}B_{1} + \frac{1}{z+y}B_{y}+\frac{1}{z-(1-y)}B_{1-y}\right)\vec{I}(y,z;\epsilon)\,,\label{eqn: DEz}
\end{align}
where $A_{i}, B_{i}$ are sparse matrices of rational numbers. It is obvious from this form that the solutions for the canonical combinations can be expressed in terms of MPLs~\cite{Goncharov:1998kja} with the alphabet 
$\{y,z,y-1,z-1,y+z,1-y-z\}$, which are usually written out in terms of a fibration in either $y$ or $z$.
We recall here that MPLs are defined as iterated integrals over rational functions
\begin{align}
G(l_1,...,l_n;x) = \int_0^x \frac{dt}{t-l_1}G(l_2,...,l_n;t)\,,
\quad G(\underbrace{0,...,0}_n;x) = \frac{1}{n!} \log^n(x)\,,
\end{align}
with $G(x) = 1$. In this context, $n$ is referred to as the \emph{transcendental weight} of the polylogarithm. 

By construction, the method of differential equations cannot be used to compute
purely one-scale integrals directly. These must instead be obtained from alternative 
methods and added to the system of differential equations.
At two loops, there are five such two- and three-point functions:
\begin{align}
I_{1} &= \epsilon^2 (1-y-z) \left[
\begin{tikzpicture}[baseline=(a1.base), scale = 2]
\begin{feynman}[inline=(a1.base)]
\vertex (x1) at (-0.5,0.25) {\(p_{1}\)};
\vertex (y1) at (-0.5,-0.25) {\(p_{2}\)};
\vertex (b1) at (-0.25,0) ;
\vertex (b3) at (0.25,0) ;
\vertex (a3) at (0.5,0) {\(p_{12}\)} ;
\diagram* {
(x1) -- [] (b1),
(y1) -- [] (b1),
(b1) -- [] (b3),
(b3) -- [ half left] (b1),
(b1) -- [ half left] (b3),
(b3) -- [scalar] (a3),
};
\vertex [dot] (i) at ($(b1)!0.5!(b3) + (0, 0.21)$) {};
\vertex [dot] (j) at ($(b1)!0.5!(b3) - (0, 0.21)$) {};
\end{feynman}
\end{tikzpicture}\right]\,,\\
       I_{2} &= -\epsilon^2 \left[
\begin{tikzpicture}[baseline=(a1.base), scale = 2]
\begin{feynman}[inline=(a1.base)]
\vertex (x1) at (-0.5,0.25) {\(p_{1}\)};
\vertex (a1) at (-0.5,0) {\(p_{2}\)};
\vertex (y1) at (-0.5,-0.25) {\(p_{3}\)};
\vertex (b1) at (-0.25,0) ;
\vertex (b3) at (0.25,0) ;
\vertex (a3) at (0.5,0) {\(p_4\)} ;
\diagram* {
(x1) -- [] (b1),
(a1) -- [] (b1),
(y1) -- [] (b1),
(b1) -- [] (b3),
(b3) -- [ half left] (b1),
(b1) -- [ half left] (b3),
(b3) -- [scalar] (a3),
};
\vertex [dot] (i) at ($(b1)!0.5!(b3) + (0, 0.21)$) {};
\vertex [dot] (j) at ($(b1)!0.5!(b3) - (0, 0.21)$) {};
\end{feynman} \label{eqn:inputfirst}
\end{tikzpicture}\right]\,,\\ 
 I_{6} &= \epsilon^3 ( 1-y-z) \left[
\begin{tikzpicture}[baseline=(a1.base), scale = 2]
\begin{feynman}[inline=(a1.base)]
\vertex (x1) at (-1,0.25) {\(p_{1}\)};
\vertex (y1) at (-1,-0.25) {\(p_{2}\)};
\vertex (b1) at (-0.5,0) ;
\vertex (b2) at (0,0.25);
\vertex (b3) at (0,-0.25);
\vertex (v2) at (0.5,0.25) {\(p_{1}\)} ;
\vertex (v3) at (0.5,-0.25) {\(p_{2}\)} ;
\diagram* {
(x1) -- [] (b1),
(y1) -- [] (b1),
(b1) -- [ ] (b2),
(b2) -- [quarter left ] (b3),
(b2) -- [ quarter right] (b3),
(b1) -- [ ] (b3),
(b3) -- [ ] (v3),
(b2) -- [] (v2),
};
\vertex [dot] (i) at ($(b2)!0.5!(b3) + (0.1, 0)$) {};
\end{feynman}
\end{tikzpicture}\right]\,,\\
 I_{9} &= \epsilon^2 ( 1-y-z)^2 \left[
\begin{tikzpicture}[baseline=(a1.base), scale = 2]
\begin{feynman}[inline=(a1.base)]
\vertex (x1) at (-1,0.25) {\(p_{1}\)};
\vertex (y1) at (-1,-0.25) {\(p_{2}\)};
\vertex (b1) at (-0.5,0) ;
\vertex (b2) at (0,0);
\vertex (b3) at (0.5,0) ;
\vertex (a3) at (1,0) {\(p_{12}\)} ;
\diagram* {
(x1) -- [] (b1),
(y1) -- [] (b1),
(b1) -- [ half left] (b2),
(b2) -- [ half left] (b1),
(b2) -- [ half left] (b3),
(b3) -- [ half left] (b2),
(b3) -- [scalar] (a3),
};
\vertex [dot] (i) at ($(b1)!0.5!(b2) + (0, 0.21)$) {};
\vertex [dot] (j) at ($(b2)!0.5!(b3) + (0, 0.21)$) {};
\end{feynman}
\end{tikzpicture}\right]\,,\\
 I_{68} &= \epsilon^4 z^2 \left[
\begin{tikzpicture}[baseline=(a1.base), scale = 2]
\begin{feynman}[inline=(a1.base)]
\vertex (x1) at (-1,0.25) {\(p_{2}\)};
\vertex (y1) at (-1,-0.25) {\(p_{3}\)};
\vertex (b1) at (-0.5,0) ;
\vertex (b2) at (0,0.25);
\vertex (b3) at (0,-0.25);
\vertex (z2) at (-0.25,0.25);
\vertex (z3) at (-0.25,-0.25);
\vertex (v2) at (0.5,0.25) {\(p_{2}\)} ;
\vertex (v3) at (0.5,-0.25) {\(p_{3}\)} ;
\diagram* {
(x1) -- [] (b1),
(y1) -- [] (b1),
(b1) -- [ ] (z2),
(z2) -- [ ] (b3),
(z3) -- [ ] (b2),
(b1) -- [ ] (z3),
(b3) -- [ ] (v3),
(b2) -- [] (v2),
(b3) -- [ ] (z3),
(b2) -- [] (z2),
};
\end{feynman}\label{eqn:inputlast}
\end{tikzpicture}\right]\,,
\end{align}
and their kinematic crossings.
Analytical expressions for all these one-scale integrals are  known in closed form in the dimensional regulator $\epsilon$~\cite{Gehrmann:1999as,Gehrmann:2005pd}.

We approach the solution of the differential equations for the remaining integrals as follows. At each order $n$ in $\epsilon$, we consider the vector of master integrals $\vec{I}^{(n)}(y,z)$ and by choice integrate first the equations in the variable $y$. If the set of differential equations in the two variables are consistent, we obtain a partial solution which differs from the full solution $\vec{I}^{(n)}(y,z)$ only by a function of the other variable, $\vec{f}(z)$,
\begin{equation}
\vec{I}^{(n)}(y,z) = \vec{I}^{(n)}_y(y,z) + \vec{f}(z)\,.
\end{equation}
The intermediate result is then substituted into the set of equations in $z$ 
\begin{align}
 \frac{\partial}{\partial z} \vec{I}^{(n)}(y,z) = \frac{\partial}{\partial z} \vec{I}^{(n)}_{y}(y,z) +  \frac{\partial}{\partial z}  \vec{f}(z) \overset{!}{=} B\vec{I}^{(n-1)}(y,z)\,
 \end{align}
and solved for $\vec{f}(z)$, which fixes the final solution up to a numerical constant
\begin{align}
\vec{I}^{(n)}(y,z) = \vec{I}^{(n)}_y(y,z) + \vec{I}^{(n)}_z(z) + \vec{c}\,,
\label{eqn:gensol}
\end{align}
with,
\begin{equation}
\vec{I}^{(n)}_z(z) = \int^z dz' \big[B\vec{I}^{(n-1)}(y,z') -  \frac{\partial}{\partial z'} \vec{I}^{(n)}_{y}(y,z') \big].
\label{eqn: eq for f}
\end{equation}
As stated above, \eqref{eqn: eq for f} cannot depend on $y$ and the spurious dependence must cancel from the right-hand side of the equation.

\subsection{Fixing boundary conditions}\label{subsec:bcs}
The remaining numerical constants $\vec{c}$ can in principle be fixed by evaluating the integrals at special kinematical points. We opted to obtain all boundary conditions without any additional computations simply by imposing
a set of regularity conditions on the general solution~\eqref{eqn:gensol}.
As suggested in~\cite{Gehrmann:2000zt,Gehrmann:2001ck}, 
a large set of boundary conditions can be obtained exploiting regularity of the master integrals at various pseudo-thresholds.
This can be achieved in practice by multiplying the differential equations \eqref{eqn: DEy} and \eqref{eqn: DEz} by those letters which correspond to pseudo-thresholds of the integrals, and taking the limit as follows:
\begin{alignat}{2}
\text{left-hand side}&: \lim_{x\to l_i}\left(\frac{\partial}{\partial x}\vec{I}\right) (x- l_i)&&= 0\,,\\
\text{right-hand side}&: \lim_{x\to l_i}\left(\epsilon A\vec{I}\right) (x-l_i) &&= \lim_{x\to l_i} \epsilon \sum_{j}A_j(x-l_i)\vec{I}= \epsilon A_i\lim_{x\to l_i}\vec{I} \,.
\end{alignat}
Requiring the right-hand side to vanish yields non-trivial relations between the integrals in the limit $x\to l_i$. Namely, if the rational factor in question appears in the homogeneous term of the differential equation for a given master integral, its value at a regular kinematic point can be typically related to other integrals in the same sector and its subtopologies. This approach is sufficient for the planar topology. 

In order to impose  these regularity conditions, we used \code{PolyLogTools}~\cite{Duhr:2019tlz} to manipulate multiple polylogarithms up to weight 5, evaluate the required limits, and perform changes in the fibration basis. Beyond weight 5, we had to carry out the required fibrations ourselves, building on the implementation of differentiation and integration of MPLs in \code{PolyLogTools}. In particular, we differentiated the integrals with respect to the variable we intend to fibrate into and obtained linear combinations of MPLs of weight 5, which can be treated with automated routines. The result can subsequently be integrated back and expressed in the required form, up to an integration constant.
All constants can be fixed by comparing the original and fibrated function at a kinematic point, and reconstructing their difference as a number of the appropriate weight using the PSLQ algorithm~\cite{pslq}. In practice, we find that our definition of the variables $y$ and $z$, \eqref{eqn:xyz}, allows us to consider just the 3 limits $y \to 1$, $y\to0 $, and $y\to -z$.

In the NPL topology, integrals possess branch points when $y \to 0$, $z \to 0$ and $1-y-z \to 0$. For this reason, the strategy outlined above can only be applied to a small number of letters and one can show that these conditions are not enough to fix all remaining constants. 
Taking inspiration from~\cite{Henn:2020lye,Wasser:2022kwg}, we also consider the singular limits (i.e. genuine thresholds of the master integrals). In terms of the Mandelstam variables, these correspond to the additional limits $s\to0$, $t\to0$ and $u \to 0$.
Crucially, our canonical basis consists of UV-finite integrals. Hence, we only need to regulate IR divergences and can assume that $\epsilon < 0$. Within this condition and keeping $\epsilon$ fixed, if a linear combination of integrals develops a singular behaviour for one of the singular limits,
this must correspond to a spurious UV-type divergence. Since no new UV divergences can appear when the kinematical invariants take special values, we may impose that such spurious divergences do not occur at the kinematic points where one of the letters vanishes. In this way, we obtain yet more equations between the boundary constants. Explicitly, the solution to the DE near the point $y\to l_i$ is 
\begin{align}
\vec{I}(y,z;\epsilon) = \exp{\{\epsilon\log{(y-l_i)}\lim_{y\to l_i}[yA_y]\}}\vec{I}\big|_{y=l_i} + \mathcal{O}(y)\,.
\end{align}

The matrix exponential contains elements with terms of the type $(y-l_i)^{a\epsilon}$. If $a > 0$, such an expression diverges in the limit $y\to l_i$. Since all our integrals must be finite at this kinematic point, the constants $\vec{I}\big|_{y=l_i}$ ought to take specific values such that these terms cancel. 

We also studied the asymptotic behaviour of our master integrals with \texttt{asy}~\cite{Jantzen:2012mw}, which is implemented in \texttt{FIESTA}~\cite{Smirnov:2015mct}, and verified the $\epsilon$-dependence in the limit $y\to l_i$. Note that in~\cite{Gehrmann:2001ck}, it was required to integrate one order in $\epsilon$ higher than necessary to enforce the non-appearance of spurious singularities at the previous order, which is no longer required. 
We checked our solutions for all top sector master integrals numerically against \texttt{pySecDec}~\cite{Borowka:2017idc} for several Euclidean points up to weight six and found perfect agreement. 

In Appendix \ref{sec: AppendixB_1L}, we present the computation of the one-loop master integrals to order $\mathcal{O}(\epsilon^4)$, which also serves as a simple example of some of the techniques discussed in this section.

\section{UV renormalisation and IR regularisation}\label{sec:renorm}
The bare helicity amplitudes \eqref{eqn:AmplExpansion} contain ultraviolet (UV) as well as infrared (IR) divergences that manifest as poles in the Laurent expansion in the dimensional regulator $\epsilon$. The former are treated in the $\overline{\text{MS}}$ scheme by expressing the amplitudes in terms of the renormalized couplings, $\alpha_{s} \equiv \alpha_{s}(\mu^2)$ and $\lambda \equiv \lambda(\mu^2)$, evaluated at the renormalization scale $\mu^2$. The resulting amplitudes still contain IR singularities, which will be cancelled analytically by those occurring in radiative processes of the same order~\cite{Kinoshita:1962, Lee:1964}. Their structure is universal and it was originally determined up to two loops by Catani~\cite{Catani:1997,Catani:1998}. These results were later systematised and extended to general processes and up to three loops in~\cite{Sterman:2002qn,Aybat:2006wq,Aybat:2006mz,Becher:2009cu,Becher:2009qa,Dixon:2009gx,Gardi:2009qi,Gardi:2009zv,Almelid:2015jia}. 

In this Section, we present the necessary steps and formulae to perform the UV renormalization and subtraction of the IR poles. This allows us to obtain the one-loop and two-loop finite remainders, which we decompose according to their colour structure. In contrast to~\cite{Gehrmann:2012}, where the IR subtraction was performed in the Catani scheme, we followed a subtraction scheme based on Soft-Collinear Effective Theory~\cite{Becher:2009cu,Becher:2009qa}, which can be more naturally extended to higher loops. In subsection \ref{sec:SCETtoCATANI} we provide conversion formulas between the two different schemes. 

\subsection{Ultraviolet renormalization}
We start by denoting all unrenormalized quantities with a superscript $U$ and then replace the bare coupling $\alpha^U$ with the renormalized strong coupling $\alpha_{s}\equiv \alpha_{s}(\mu^2)$, evaluated at the renormalization scale $\mu^2$,
\begin{equation}
    \alpha^U \mu_{0}^{2\epsilon}S_{\epsilon} = \alpha_s \mu^{2\epsilon} \bigg[1-\frac{\beta_0}{\epsilon}\bigg(\frac{\alpha_s}{2 \pi}\bigg)+\bigg(\frac{\beta_0^2}{\epsilon^2}-\frac{\beta_1}{2 \epsilon}\bigg)\bigg(\frac{\alpha_s}{2 \pi}\bigg)^2+\mathcal{O}(\alpha_s^3)\bigg],
\end{equation}
where $S_{\epsilon} = (4 \pi)^{\epsilon}\mathrm{e}^{-\epsilon \gamma_{E}}$ and $\mu_0^2$ is the mass parameter in dimensional regularization introduced to maintain a dimensionless coupling in the bare QCD Lagrangian density. The explicit form of the first two $\beta$-function coefficients $\beta_0,\, \beta_1$ reads 
\begin{align}
    \beta_0 &= \frac{11 C_A}{6} - \frac{2 T_R N_F}{3}, \\
    \beta_1 &= \frac{17 C_A^2}{6} - \frac{5 C_A T_R N_F}{3} -  C_F T_R N_F,
\end{align}
with the QCD colour factors,
\begin{equation}
    C_A = N,\quad C_F = \frac{N^2-1}{2N},\quad T_R = \frac{1}{2}.
\end{equation}

The effective coupling $\lambda$ is renormalized as follows,
\begin{equation}
    \lambda^U = \lambda \bigg[1-\frac{\beta_0}{\epsilon}\bigg(\frac{\alpha_S}{2 \pi}\bigg)+\bigg(\frac{\beta_0^2}{\epsilon^2}-\frac{\beta_1}{ \epsilon}\bigg)\bigg(\frac{\alpha_S}{2 \pi}\bigg)^2+\mathcal{O}(\alpha_S^3)\bigg].
\end{equation}
The renormalized coefficients of the UV-finite but IR-divergent amplitudes can be written in terms of the $i$-loop contribution to the unrenormalized coefficients $\Omega^{(i),\, U}$ as
\begin{align}
    \Omega^{(0)} &= \Omega^{(0),\, U},\\
    \Omega^{(1)} &= S_{\epsilon}^{-1}\Omega^{(1),\, U} - \frac{3\beta_0}{2\epsilon}\Omega^{(0),\, U},\\
    \Omega^{(2)} &= S_{\epsilon}^{-2}\Omega^{(2),\, U} - \frac{5\beta_0}{2\epsilon}S_{\epsilon}^{-1}\Omega^{(1),\, U} - \bigg(\frac{5\beta_1}{4\epsilon} - \frac{15\beta_0^2}{8\epsilon^2}\bigg)\Omega^{(0),\, U}.
    \label{eqn: renormHelicityAmplitude}
\end{align}
For the remainder of the paper, we will set $\mu^2 = \mu_0^2 = M_H^2$ for simplicity.

\subsection{Infrared factorization}
Since the IR poles of $l$-loop amplitudes in gauge theories factorize in colour-space in terms of lower loop amplitudes, the IR poles can be subtracted multiplicatively as
\begin{equation}
   \mathbf{\Omega}({p}, \epsilon) = \mathbfcal{Z}(\{p\}; \epsilon) \,\,\mathbf{\Omega}^{\text{finite}}(\{p\}) .
    \label{eqn: IR-Renormalization}
\end{equation}
Contrary to multiplicative renormalization of UV divergences, the bold notation in \eqref{eqn: IR-Renormalization} indicates that, in general,  $\mathbfcal{Z}$ and $\mathbf{\Omega}$ are operators and vectors in colour space, respectively. The all-order nature of \eqref{eqn: IR-Renormalization} makes this approach particularly advantageous for generalizations to higher orders in perturbation theory.

Solving a renormalization group equation for $\mathbfcal{Z}$, one finds
\begin{equation}
    \mathbfcal{Z}(\epsilon, \{p\}, \mu) = \mathbb{P} \,  \text{exp} \Big[  \int_{\mu}^{\infty} \frac{d\mu'}{\mu'} \mathbf{\Gamma}(\{p\}, \mu') \Big] = \sum_{l = 0}^{\infty} \Big( \frac{\alpha_s}{2 \pi} \Big)^l \mathbfcal{Z}^{(l)} ,
    \label{Z_operator}
\end{equation}
where $\mathbb{P}$ is the path-ordering symbol, meaning that the colour operators are ordered from left to right in decreasing values of $\mu'$. 
As originally proposed by Catani~\cite{Catani:1998}, the anomalous-dimension matrix $\mathbf{\Gamma}$ for amplitudes with $n$ QCD partons up to two loops  is entirely governed 
by the dipole colour correlations operator
\begin{equation}
\mathbf{\Gamma}(\{p\}, \mu) =  \mathbf{\Gamma}_{dipole}(\{p\},\mu),
\label{Dipole}
\end{equation}
\begin{equation}
 \mathbf{\Gamma}_{dipole}(\{p\},\mu) = \sum_{1 \leq i < j \leq n} \mathbf{T}_i^a\mathbf{T}_j^a \, \gamma^{\text{cusp}}(\alpha_s) \log\Big(-\frac{\mu^2}{s_{ij} + i \eta }\Big) \, +  \, \sum_{i=1}^n \gamma^i(\alpha_s),
 \label{DipoleExpression}
\end{equation}
where $\gamma^{\text{cusp}}$ is the cusp anomalous dimension and $\gamma^i$ is the anomalous dimension of the $i$-th external particle, the latter depending on the nature of the particle. The cusp anomalous dimension carries the information about overlapping soft and collinear divergences, while $\gamma^i$ only involves collinear divergences associated to
the $i$-th parton.
The perturbative expansions up to two loops for the anomalous dimensions are listed in Appendix \ref{sec: AppendixA_IR}. The coupling constant is evaluated at the renormalization scale $\alpha_s = \alpha_s(\mu)$.

The colour operators $\mathbf{T}^a_i$ are related to the $SU(N)$ generators and their action on the $i$-th coloured parton is defined following the convention in~\cite{Catani:1998} as:
\begin{align} \label{eqn: ToperatorAction}
(\mathbf{T}_i^{a})_{b_i c_i}  &= i \, f^a_{b_ic_i} \quad \text{for a gluon}\,, \nonumber \\
(\mathbf{T}_i^{a})_{l_i k_i}  &= +T^a_{l_ik_i} \quad \text{for a final(initial) state quark(anti-quark)}\,, \\
(\mathbf{T}_i^{a})_{l_i k_i}  &= -T^a_{k_il_i} \quad \text{for a initial(final) state quark(anti-quark)} \nonumber\,.
\end{align}
where $i$ labels the particle on which the operator is acting. Importantly, colour conservation can be rephrased as
\begin{align}
\sum_i \mathbf{T}_i^a = 0.
\label{eqn: ColorConservation}
\end{align}
It follows from these definitions that the repeated action of one operator evaluates to a Casimir,
\begin{equation}
(\mathbf{T}_i^a)^2 = C_i \, ,
\label{eqn: Tsquared}
\end{equation}
 where $C_i = C_A$ if particle $i$ is a gluon and $C_i = C_F$ in case of (anti-)quark. 

We also define the expansions,
\begin{equation}
\mathbf{\Gamma}_{dipole} = \sum_{l=0}^{\infty} \mathbf{\Gamma}_{l} \, \,  \Big(\frac{\alpha_s}{2 \pi} \Big)^{l+1}, \quad \quad \, \Gamma' = \frac{\partial \mathbf{\Gamma}_{dipole}}{\partial \log(\mu)} = \sum_{l=0}^{\infty} \Gamma'_{l} \, \,  \Big(\frac{\alpha_s}{2 \pi} \Big)^{l+1},
\label{DipoleExpansion}
\end{equation}
where one can drop the bold notation in the derivative because the resulting operator is always diagonal in colour space:
\begin{align}
\Gamma' = - \gamma^{\text{cusp}}(\alpha_s) \sum_i C_i.
\label{eqn: gamma prime}
\end{align}

In our specific case, expanding \eqref{eqn: IR-Renormalization} up to two loops, 
IR divergences in the renormalized two loop amplitudes can be expressed in terms of the renormalized tree and one-loop amplitudes multiplied by appropriate operators
\begin{align}
\mathbf{\Omega}^{(0)}_{\text{finite}} & = \mathbf{\Omega}^{(0)},\\
\mathbf{\Omega}^{(1)}_{\text{finite}} & = \mathbf{\Omega}^{(1)} - \mathbf{I}_{\Omega}^{(1)} \, \mathbf{\Omega}^{(0)} ,\label{eqn: perturbative IR subtraction one loop}\\
\mathbf{\Omega}^{(2)}_{\text{finite}} & = \mathbf{\Omega}^{(2)} - \mathbf{I}_{\Omega}^{(2)} \, \mathbf{\Omega}^{(0)}  - \mathbf{I}_{\Omega}^{(1)} \, \mathbf{\Omega}^{(1)},
\label{eqn: perturbative IR subtraction two loops}
\end{align}
With the definitions given in \eqref{Z_operator}--\eqref{DipoleExpansion}, we can express the subtraction operators in \eqref{eqn: perturbative IR subtraction one loop}--\eqref{eqn: perturbative IR subtraction two loops} as
\begin{align}
 \mathbf{I}^{\text{SCET},(1)} & = \mathbfcal{Z}^{(1)},\\
 \mathbf{I}^{\text{SCET},(2)} & = \mathbfcal{Z}^{(2)} - \big(\mathbfcal{Z}^{(1)}\big)^2,
 \label{eqn: IR scet}
\end{align}
where the process dependence is implicit and
\begin{align}
 \mathbfcal{Z}^{(1)} &= \frac{\Gamma'_0}{4 \epsilon^2} + \frac{\mathbf{\Gamma_0}}{2 \epsilon},\\
\mathbfcal{Z}^{(2)} &= \frac{{\Gamma_0'}^2}{32 \epsilon^4} +  \frac{\Gamma'_0}{8 \epsilon^3}\Big(\mathbf{\Gamma_0}- \frac{3}{2}\beta_0 \Big) + \frac{\mathbf{\Gamma_0}}{8 \epsilon^2}\Big( \mathbf{\Gamma}_0 - 2 \beta_0\Big) + \frac{\Gamma'_1}{16 \epsilon^2} + \frac{\mathbf{\Gamma}_1}{4 \epsilon}.
\label{eqn: Z scet}
\end{align}
It is now manifest that the IR operators in SCET scheme have only pole terms, without any $\mathcal{O}(\epsilon^0)$ contributions.

 The cancellation of IR poles according to \eqref{eqn: perturbative IR subtraction two loops} is a strong analytic check on multiloop amplitudes. We shall now follow the SCET approach to derive explicit expressions for the subtraction operators.

\subsection{SCET operators for three coloured partons: \( ggg\) and $q\bar{q}g$}
The definition of the finite parts of a two-loop amplitude is based on \eqref{eqn: IR scet}, which we now  work out for our process.
We start from $\Gamma'_{n}$, which can be read off directly from \eqref{eqn: gamma prime} and the dependence on the nature of the three partons is entirely contained in the sum over the Casimirs.
On the other hand, for the $\mathbf{\Gamma}_{n}$ coefficients, we can first notice that an additional simplification occurs when only three coloured partons are involved. 
In this case, the resulting amplitude is proportional to one single colour structure, namely $f^{abc}$ is the colour factor for $H \rightarrow g(a) + g(b) + g(c)$, while $T^a_{ij}$ is the factor for $H \rightarrow q(i) + \bar{q}(j) + g(c)$.

As a consequence~\cite{Catani:1998}, the dipole operator itself diagonalizes in colour space. In fact, using colour conservation \eqref{eqn: ColorConservation} and the property \eqref{eqn: Tsquared}, one can rewrite the product of two $\mathbf{T}^a$ operators as a sum over Casimirs
\begin{align}
2 \, \mathbf{T}_1^a \mathbf{T}_2^a &= (\mathbf{T}_3^a )^2 - (\mathbf{T}_1^a )^2 -(\mathbf{T}_1^a )^2 = C_3 - C_1 - C_2\,, \nonumber\\
2 \, \mathbf{T}_2^a \mathbf{T}_3^a &= (\mathbf{T}_1^a )^2 - (\mathbf{T}_2^a )^2 -(\mathbf{T}_3^a )^2 = C_1 - C_2 - C_3\,, \nonumber \\
2 \, \mathbf{T}_1^a \mathbf{T}_3^a &= (\mathbf{T}_2^a )^2 - (\mathbf{T}_1^a )^2 -(\mathbf{T}_3^a )^2 = C_2 - C_1 - C_3.
\end{align}
Accordingly, we can drop the bold notation and we find
expressions relevant to the three helicity 
coefficients $\alpha,\beta,\gamma$:
\begin{align}
\Gamma^{\alpha}_{n} &=  \Gamma^{\beta}_{n} =- \frac{C_A}{2} \Big( L_{12} + L_{23} + L_{13} \Big) \gamma^{\text{cusp}}_n  + 3 \gamma_n^{g}, \\
\Gamma^{\gamma}_{n} &=  -C_F \, L_{12} \,  \gamma^{\text{cusp}}_n - \frac{C_A}{2} \Big(-L_{12} + L_{23} + L_{13} \Big) \gamma^{\text{cusp}}_n  + 2 \gamma_n^{q} + \gamma_n^{g},
\label{eqn: GammaExpansion}
\end{align}
with 
\begin{align}
L_{ij} = \log{\Big(-\frac{\mu^2}{s_{ij}+ i \eta} \Big)}
\end{align}
and the anomalous dimension coefficients defined as in Appendix \ref{sec: AppendixA_IR}.

We stress here that there is an imaginary part arising from the logarithms above whenever the corresponding invariant is positive, which depends on the kinematical region considered. 

\subsection{Results for the helicity amplitudes}
We computed the renormalized amplitudes for the decay process up to order $\epsilon^4$ at one loop and up to order $\epsilon^2$ at two loops. Working in the SCET subtraction scheme, we derived the finite remainder for all the helicity amplitudes.
Decomposed according to their colour structure, they read
\begin{align}
\Omega^{(1)}_{{\rm finite}} &= \bigg(N\ A^{(1)}_\Omega + \frac{1}{N}\ B^{(1)}_\Omega + N_F C^{(1)}_\Omega \bigg), \label{eqn: renormalized1}\\
\Omega^{(2)}_{{\rm finite}} &=  \bigg(N^2 A^{(2)}_\Omega +N^0 B^{(2)}_\Omega +\frac{1}{N^2} C^{(2)}_\Omega +\frac{N_F}{N} D^{(2)}_\Omega  +  N N_F E^{(2)}_\Omega + N_F^2 F^{(2)}_\Omega \bigg). \label{eqn: renormalized2} 
\end{align} 
The same structure holds for the renormalized amplitudes, though the respective coefficients will still contain poles in $\epsilon$.

In the supplementary material, we provide the coefficients of the renormalized amplitudes for the decay processes and of the finite remainder for the Higgs decay kinematics as well as for all crossings~\cite{Gehrmann:2002zr} relevant to 
$H+$jet production processes.

\subsection{Conversion to the Catani scheme}\label{sec:SCETtoCATANI}
The two-loop helicity amplitudes for the decay of a Higgs boson into three partons were first computed in~\cite{Gehrmann:2012}. The authors obtained the finite remainder by subtracting the IR singularities according to the 
original Catani prescription~\cite{Catani:1998}. Here we give the conversion rules between the two subtraction schemes, which also served as a cross-check of our results. 

Following closely the notation of~\cite{Gehrmann:2012}, we write the subtraction operators $\mathbf{I}_{\Omega}$ of \eqref{eqn: perturbative IR subtraction one loop} in Catani scheme as
\begin{align}
    \mathbf{I}_{\alpha}^{\text{C},(1)} &= \mathbf{I}_{\beta}^{\text{C},(1)} =  - \frac{\mathrm{e}^{\epsilon \gamma}}{2 \Gamma(1-\epsilon)} \bigg[N\bigg(\frac{1}{\epsilon^2} + \frac{\beta_0}{N \epsilon}\bigg)\big({\tt S}_{12} + {\tt S}_{23} + {\tt S}_{13}\big)\bigg], \nonumber \\
    \mathbf{I}_{\gamma}^{\text{C},(1)} &= - \frac{\mathrm{e}^{\epsilon \gamma}}{2 \Gamma(1-\epsilon)}\bigg[N\bigg(\frac{1}{\epsilon^2} + \frac{3}{4\epsilon} + \frac{\beta_0}{2 N \epsilon}\bigg)\big({\tt S}_{23} + {\tt S}_{13}\big) - \frac{1}{N}\bigg(\frac{1}{\epsilon^2} + \frac{3}{2\epsilon}\bigg){\tt S}_{12}\bigg], 
\end{align}
with,
\begin{equation}
    {\tt S}_{ij} = \bigg(- \frac{\mu^2}{s_{ij}}\bigg)^\epsilon.
    \label{eqn: Sij catani}
\end{equation}
The second-order operator can be built starting from the one-loop operator as
\begin{align}
\mathbf{I}_{\Omega}^{\text{C},(2)} =& -\frac{1}{2} \mathbf{I}^{\text{C},(1)}_{\Omega}(\epsilon) \, \mathbf{I}^{\text{C},(1)}_{\Omega}(\epsilon) - \frac{\beta_0}{\epsilon}\mathbf{I}^{\text{C},(1)}_{\Omega}(\epsilon) \nonumber \\ & + \mathrm{e}^{-\epsilon \gamma} \frac{\Gamma(1-2 \epsilon)}{\Gamma(1-\epsilon)} \bigg(\frac{\beta_0}{\epsilon} + K\bigg) \mathbf{I}^{\text{C},(1)}_{\Omega}(2 \epsilon) + \mathbf{H}^{(2)}_{\Omega}(\epsilon)\,, 
\label{eqn: IR2 catani}
\end{align}
where we introduced the constant
\begin{equation}
    K = \bigg(\frac{67}{18} - \frac{\pi^2}{6}\bigg)C_A - \frac{10}{9}T_R N_F.
\end{equation}

The remaining term in \eqref{eqn: IR2 catani} involves the operator $\mathbf{H}^{(2)}_{\Omega}(\epsilon)$ and produces only a single pole in $\epsilon$. Its explicit form is
\begin{equation}
    \mathbf{H}^{(2)}_{\Omega}(\epsilon) = \frac{\mathrm{e}^{\epsilon \gamma}}{4 \epsilon \Gamma(1-\epsilon)}  H^{(2)}_{\Omega}.
\end{equation}
The constant $H^{(2)}_{\Omega}$ is renormalization scheme and process dependent and in our case it reads
\begin{align}
    H^{(2)}_{\alpha} &= H^{(2)}_{\beta} = 3 H^{(2)}_{g},\\
    H^{(2)}_{\gamma} &= 2 H^{(2)}_{q} + H^{(2)}_{g},
\end{align}
where in the $\overline{\text{MS}}$ scheme the constants $H^{(2)}_{q},\, H^{(2)}_{g}$ are
\begin{align}
    H^{(2)}_q &=
\left({7\over 4}\zeta_3+{\frac {409}{864}}- {\frac {11\pi^2}{96}}
\right)N^2
+\left(-{1\over 4}\zeta_3-{41\over 108}-{\pi^2\over 96}\right)
+\left(-{3\over 2}\zeta_3-{3\over 32}+{\pi^2\over 8}\right){1\over
N^2}\nonumber \\
&
+\left({\pi^2\over 48}-{25\over 216}\right){(N^2-1)N_F\over N}\;, \\
H^{(2)}_g &=  
\left(\frac{1}{2}\zeta_3+{\frac {5}{12}}+ {\frac {11\pi^2}{144}}
\right)N^2
+{\frac {5}{27}}\, N_F^2
+\left (-{\frac {{\pi }^{2}}{72}}-{\frac {89}{108}}\right ) N N_F 
-\frac{N_F}{4N}. 
\end{align}

When describing the SCET subtraction scheme, we pointed out that the subtraction operators have no finite $\mathcal{O}(\epsilon^0)$ contribution.

In contrast, the Catani operators contain coefficients at higher order in the dimensional regulator, generated by the $\epsilon$ expansion of the resummed coefficient defined in Eq. \eqref{eqn: Sij catani}.
The tree-level amplitude is finite. One-loop amplitudes have poles starting from order $\mathcal{O}(1/\epsilon^{2})$ and two-loop amplitudes have poles starting from order $\mathcal{O}(1/\epsilon^4)$. We indicate with $\mathbf{\Omega}_{n}^{(l)}$, $\mathbf{I}_{n}^{\text{SCET},(l)}$ and  $\mathbf{I}_{n}^{\text{C},(l)}$ the coefficients of order $\epsilon^n$ of the renormalized amplitude, the SCET operator and the Catani operator, respectively. 

From the subtraction formulae \eqref{eqn: perturbative IR subtraction one loop}--\eqref{eqn: perturbative IR subtraction two loops}, it is easy to obtain 
the following conversion rules for the finite remainders
\begin{align}
\mathbf{\Omega}_{\text{finite}}^{\text{SCET},(0)} - \mathbf{\Omega}_{\text{finite}}^{\text{C},(0)} &= 0, \\
\mathbf{\Omega}_{\text{finite}}^{\text{SCET},(1)} - \mathbf{\Omega}_{\text{finite}}^{\text{C},(1)}  &=  \mathbf{I}_{0}^{\text{C},(1)} \mathbf{\Omega}^{(0)} ,  \\
\mathbf{\Omega}_{\text{finite}}^{\text{SCET},(2)} - \mathbf{\Omega}_{\text{finite}}^{\text{C},(2)} &=   \mathbf{I}_{0}^{\text{C},(2)} \mathbf{\Omega}^{(0)} 
+ \mathbf{I}_{2}^{\text{C},(1)} \mathbf{\Omega}_{-2}^{(1)} 
+\mathbf{I}_{1}^{\text{C},(1)} \mathbf{\Omega}_{-1}^{(1)} 
+ \mathbf{I}_{0}^{\text{C},(1)} \mathbf{\Omega}_{0}^{(1)}.
\label{eqn: finite difference scheme}
\end{align}
In deriving the above rules, we made use of the fact that $\mathbf{I}^{\text{C},(1)}$ and $\mathbf{I}^{\text{SCET},(1)}$ have the same pole structure. Consequently, terms multiplying $(\mathbf{I}^{\text{C},(1)}_{-1}-\mathbf{I}^{\text{SCET},(1)}_{-1})$ or $(\mathbf{I}^{\text{C},(1)}_{-2}-\mathbf{I}^{\text{SCET},(1)}_{-2})$ in \eqref{eqn: finite difference scheme} vanish. This can be easily understood by inspecting the origin of the poles in the one-loop cancellation in \eqref{eqn: perturbative IR subtraction one loop}. 

For both the decay and the production kinematics, we verified that the result given in~\cite{Gehrmann:2012} is correctly reproduced converting our finite remainder with the above rules.

\section{Analytic Continuation}\label{sec:ancont}
So far, we have described our calculation making explicit reference to decay processes, for which all the kinematic invariants are positive. We considered the decay of a Higgs boson into three gluons, $H \rightarrow ggg$, and into a quark-antiquark pair and gluon, $H \rightarrow q\bar{q} g$.
In view of applications to LHC physics, 
we are interested in the production regions, in which a Higgs is produced together with a parton: $gg \rightarrow Hg$, $ q g \rightarrow H q$, $\bar{q} g \rightarrow H \bar{q}$ and $q \bar{q} \rightarrow H g $. 

The general strategy for performing the analytic continuation for the MPLs appearing in $2\to 2$ scattering involving 4-point functions with one external off-shell leg and massless propagators was outlined in  detail in~\cite{Gehrmann:2002zr}. Our aim is to describe how this strategy can be applied to the process at hand. Referring to Fig.~1 of reference~\cite{Gehrmann:2002zr} and using the same labels for the various kinematic regions, the goal is to find a procedure to analytically continue MPLs from the decay region (1a) to the three production regions, (2a), (3a), (4a). Whenever a particle is crossed from the initial state to the final state (or vice versa), two of the invariants become negative and one remains positive, representing the centre-of-mass energy of the incoming partons. 
We recall here that results in the decay region (1a) are expressed as MPLs of the variables $y$ and $z$ defined in \eqref{eqn:xyz}, which fulfil the constraints $ 0 <  z < 1 \, , \, \, 0 < y < 1-z$. Within these bounds, our set of MPLs are real and no branching point is crossed. 

To describe the analytic continuation to the scattering kinematics, let us consider the case of region (3a). In (3a), the kinematic constraints are $ z < 0 \, , \, \, 1-z < y < +\infty $ and MPLs must be evaluated across a branch cut, developing an imaginary part. Physically, the particle with momentum $p_2$ is crossed and the process is $p_1 + p_3 \rightarrow p_2 + p_4$. Crucially, there exists a change of variables which maps region (3a) \emph{linearly} back into the decay region (1a) and can be implemented analytically on our MPLs. 
In fact, by defining
\begin{align}
 v \equiv 1 / y\,, \quad \quad \quad u \equiv - z / y\,,
 \label{eqn: uv}
\end{align}
the new variables $u$ and $v$ satisfy again $ 0 < u < 1\, , \, \, 0 < v < 1-u$. By re-expressing the MPLs in terms of $u$ and $v$, the imaginary part can be made explicit in terms of multiple zeta values and real-valued MPLs. The same manipulations can be performed in regions (2a) and (4a) with different definitions of the $(u,v)$ variables. In general, the $v$ variable is the reciprocal of the centre-of-mass energy and so its definition depends on which particle is crossed from the final to the initial state. 

Instead of performing the analytic continuation in all the three regions (2a), (3a) and  (4a), we found it simpler to first consider suitable crossings of the amplitude in the decay region and in a second step continue these crossed amplitudes only to the region (3a).
Explicitly, for the case of three gluons, the two independent amplitudes in the decay kinematics are
\begin{align}
\mathcal{M}^{+++} : \quad \quad  H &\rightarrow g_+(p_1) + g_+(p_2) + g_+(p_3)\,,  \\
\mathcal{M}^{++-} : \quad \quad H &\rightarrow g_+(p_1) + g_+(p_2) + g_-(p_3)\,.
\end{align}
In the production region, there are eight different helicity configurations. Thanks to parity symmetry, we can limit ourselves to consider just four of them (see the left column of Table~\ref{tab:crossings}) and relate them to the other four (right column).
\begin{table}[t]
\begin{center}
\renewcommand{\arraystretch}{1.2}
\begin{tabular}{c c}
   \multicolumn{1}{c}{Process} & \multicolumn{1}{c}{Parity Related} \\
   \hline
  $\mathcal{M}^{--,+}_{gg,g}: \, \, g_-(p_1) + g_-(p_3) \rightarrow H + g_+(p_2)\, \quad $    &   $\quad \quad \,g_+(p_1) + g_+(p_3) \rightarrow H + g_-(p_2)$, \\
   $\mathcal{M}^{-+,+}_{gg,g}: \, \, g_-(p_1) + g_+(p_3) \rightarrow H + g_+(p_2)\, \quad$    &   $\quad \quad \,g_+(p_1) + g_-(p_3) \rightarrow H + g_-(p_2)$, \\
  $\mathcal{M}^{--,-}_{gg,g}: \, \, g_-(p_1) + g_-(p_3) \rightarrow H + g_-(p_2)\, \quad$    &   $\quad \quad \,g_+(p_1) + g_+(p_3) \rightarrow H + g_+(p_2)$, \\
   $ \mathcal{M}^{+-,+}_{gg,g}: \, \, g_+(p_1) + g_-(p_3) \rightarrow H + g_+(p_2)\, \quad$    &   $\quad \quad \,g_-(p_1) + g_+(p_3) \rightarrow H + g_-(p_2)$. \\ 
\end{tabular}
\end{center}
\caption{Different kinematical crossings of the $H\to ggg$ amplitudes.  The comma is used to separate the helicities of initial and final state partons.
\label{tab:crossings}}
\end{table}
 Note that in continuing to the region (3a), the momenta $p_1$ and $p_3$ are always in the initial state. We computed the first three amplitudes with a combination of crossings and analytic continuation as follows (an extra helicity flip due to time reversal is always understood after continuation to (3a)): 
\begin{align}
\mathcal{M}^{+++}_{ggg}\, &\xrightarrow[]{\text{to (3a)}} \,  \mathcal{M}^{--,+}_{gg,g} \,,\\
\mathcal{M}^{++-}_{ggg}\, &\xrightarrow[]{\text{to (3a)}} \,  \mathcal{M}^{-+,+}_{gg,g} \,,\\
\mathcal{M}^{++-}_{ggg}\, \xrightarrow[]{p_2 \leftrightarrow p_3} \,& \mathcal{M}^{+-+}_{ggg}\, \xrightarrow[]{\text{to (3a)}} \,  \mathcal{M}^{--,-}_{gg,g}\,.
\end{align}
The fourth amplitude, $\mathcal{M}^{+-,+}_{gg,g}$, can be derived from $\mathcal{M}^{-+,+}_{gg,g}$ by the crossing $p_1 \leftrightarrow p_3$, which implies $v \rightarrow v$ and $ u \rightarrow 1 -u -v$. Since no branch cut is crossed under this transformation, no new analytic continuation is needed and we can conclude that
\begin{align}
\mathcal{M}^{+-,+}_{gg,g}(v,u) =  \mathcal{M}^{-+,+}_{gg,g}(v,u)\vert_{u \rightarrow 1-u-v}\,.
\end{align}

Let us consider now the decay into a quark-antiquark pair and a gluon:
\begin{align}
\mathcal{M}^{LR+} : \quad \quad  H &\rightarrow q_L(p_1) + \bar{q}_R(p_2) + g_+(p_3)\,.
\end{align}
In the production region there are 12 non-zero independent helicity configurations, but only 3 of them need to be computed, while the others are related by parity and charge conjugation. Our choice of the independent configurations is according to the 
left column of Table~\ref{tab:crossQ}.

\begin{table}[t]
\begin{center}
   \renewcommand{\arraystretch}{1.2}
\begin{tabular}{c c}
   \multicolumn{1}{c}{Process} & \multicolumn{1}{c}{C, P, CP} \\
   \hline
  $\mathcal{M}^{R-,R}_{qg,q}: \, \, q_R(p_1) + g_-(p_3) \rightarrow H + q_R(p_2)\, \quad $    &   $\quad \quad \,\bar{q}_L(p_1) + g_+(p_3) \rightarrow H + \bar{q}_L(p_2)$ \\
  $\,$ & $\quad \quad \, q_L(p_1) + g_+(p_3) \rightarrow H + q_L(p_2)$ \\ 
   $\,$ & $\quad \quad \, \bar{q}_R(p_1) + g_-(p_3) \rightarrow H + \bar{q}_R(p_2)$  \\
$\mathcal{M}^{L-,L}_{\bar{q}g,\bar{q}}: \, \, \bar{q}_L(p_1) + g_-(p_3) \rightarrow H + \bar{q}_L(p_2)\, \quad$    &   $\quad \quad \,q_R(p_1) + g_+(p_3) \rightarrow H + q_R(p_2)$ \\
  $\, $ & $\quad \quad \, \bar{q}_R(p_1) + g_+(p_3) \rightarrow H + \bar{q}_R(p_2)$ \\ 
   $\,$ & $\quad \quad \, q_L(p_1) + g_-(p_3) \rightarrow H + q_L(p_2)$ \\
  $\mathcal{M}^{RL,+}_{q\bar{q},g}: \, \, q_R(p_1) + \bar{q}_L(p_3) \rightarrow H + g_+(p_2)\, \quad $    &   $\quad \quad \,q_R(p_1) + \bar{q}_L(p_3) \rightarrow H + g_-(p_2)$ \\
  $\,$ & $\quad \quad \, q_L(p_1) + \bar{q}_R(p_3) \rightarrow H + g_-(p_2)$ \\ 
   $\,$ & $\quad \quad \, \bar{q}_R(p_1) + q_L(p_3) \rightarrow H + g_+(p_2)$ \\ 
\end{tabular}
\end{center}
\caption{Different kinematic crossings of the $H\to q\bar q g$ amplitudes. The comma is used to separate the helicities of initial and final state partons. \label{tab:crossQ}}
\end{table}
The three amplitudes were computed with a combination of crossings and analytic continuation as follows (an additional helicity flip due to time reversal is again understood after continuation to (3a)),
\begin{align}
\mathcal{M}^{LR+}_{q\bar{q}g}\, &\xrightarrow[]{\text{to (3a)}} \,  \mathcal{M}^{R-,R}_{qg,q} \\ 
\mathcal{M}^{LR+}_{q\bar{q}g}\, \xrightarrow[]{p_1 \leftrightarrow p_2} \,&  \mathcal{M}^{RL+}_{\bar{q}qg}\, \xrightarrow[]{\text{to (3a)}} \,  \mathcal{M}^{L-,L}_{\bar{q}g,\bar{q}}  \\ 
\mathcal{M}^{LR+}_{q\bar{q}g}\, \xrightarrow[]{p_2 \leftrightarrow p_3} \,&  \mathcal{M}^{L+R}_{qg\bar{q}}\, \xrightarrow[]{\text{to (3a)}} \,  \mathcal{M}^{RL,+}_{q\bar{q},g} \,.
\end{align}
Finally, we stress that each crossing must also be applied to the spinor prefactors in \eqref{eqn:HA_ggg}. We have cross-checked all amplitudes in all helicity configurations against previously published results \cite{Gehrmann:2012} at the level of the finite remainder up to $\mathcal{O}(\epsilon^0)$ and found perfect agreement. Conventions and normalization of helicity amplitudes have been fixed by numerical evaluation of our result, in the $q \bar{q} g$ channel, using \code{OpenLoops 2}~\cite{Buccioni:2019sur}.

\section{Conclusions}\label{sec:conclusions}
In this paper, we presented the calculation of the two loop corrections to the helicity amplitudes for the $H\to ggg$ and $H\to q\bar{q}g$ up to order $\mathcal{O}(\epsilon^2)$ in the large top Higgs effective field theory. 
These amplitudes constitute the first missing ingredient towards the calculation of $H+$jet production to N$^3$LO at the LHC, as they are required to properly define the finite remainder of the corresponding three loop virtual corrections. 
We follow a standard approach to compute the helicity amplitudes. We start by decomposing the amplitude in a basis of
independent tensor structures, and we use spinor-helicity to express the helicity amplitudes in terms of linear combinations
of the corresponding scalar form factors.
Next, we derived a canonical basis for the relevant master integrals and, through the method of differential equations, provided their solution up to weight six in MPLs. We verified that our results for the finite remainder of the amplitude match the literature~\cite{Gehrmann:2012}, and analytically continued the helicity amplitudes to all kinematic regions. The infrared structure of the result was inspected in the frameworks of SCET and the Catani infrared factorization formula. This updated result marks the first step towards computing the N$^3$LO corrections to $H$+jet production.

\section*{Acknowledgments}
This work was supported in part by the Excellence Cluster ORIGINS funded by the Deutsche Forschungsgemeinschaft (DFG, German Research Foundation) under Germany’s Excellence Strategy – EXC-2094-390783311, by the Swiss National Science Foundation (SNF) under contract 200020-204200, and by the European Research Council (ERC) under the European Union’s research and innovation programme grant agreements 949279 (ERC Starting Grant HighPHun) and 101019620 (ERC Advanced Grant TOPUP).

\appendix

\section{Anomalous dimensions}\label{sec: AppendixA_IR}
In this appendix, we give the perturbative coefficients for the cusp anomalous dimension $\gamma^{\text{cusp}}$, the quark anomalous dimension $\gamma_{q}$ and the gluon anomalous dimension $\gamma_g$ up to two loops, $\mathcal{O}(\alpha_s^2)$. The perturbative expansion reads
\begin{align}
\gamma^{\text{cusp}} = \sum_{n = 0}^{\infty} \gamma_n^{\text{cusp}} \Big( \frac{\alpha_s}{2 \pi} \Big)^{n+1} \, \quad \, \gamma^{i} = \sum_{n = 0}^{\infty} \gamma_n^{i} \Big( \frac{\alpha_s}{2 \pi} \Big)^{n+1}
\end{align}
with $i=q,\overline{q},g$. \\
The cusp anomalous dimension was computed at two loops in~\cite{korchemksy:1987}
\begin{align}
    \gamma_0^{\text{cusp}} &= 2, \nonumber \\
    \gamma_1^{\text{cusp}} &= \Big( \frac{67}{9} - \frac{\pi^2}{3} \Big) C_A - \frac{10}{9}N_f\,, 
\end{align}
while for the (anti-)quark anomalous dimension we have~\cite{vanneerven:1986,matsuura:1988sm},
\begin{align}
    \gamma_0^{q} &= -\frac{3}{2} \, C_F, \nonumber \\
    \gamma_1^{q} &= C_F^2 \Big(  -\frac{3}{8} + \frac{\pi^2}{2}  - 6 \zeta_3\Big) +
    C_F C_A \Big(- \frac{961}{216} - \frac{11}{24} \pi^2  + \frac{13}{2}  \zeta_3  \Big) + 
    C_F N_F \Big(\frac{65}{108} + \frac{\pi^2}{12} \Big)\,, 
\end{align}
and finally, for the gluon,~\cite{Harlander:2000},
\begin{align}
\gamma_0^{g} &= - \beta_0, \nonumber \\
\gamma_1^{g} &= C_A^2 \Big( -\frac{173}{27} + \frac{11}{72}\pi^2 + \frac{\zeta_3}{2} \Big) + C_A N_F \Big(\frac{32}{27} - \frac{\pi^2}{36}\Big) + \frac{1}{2} C_F N_F.
\end{align}

\section{One-loop master integrals}\label{sec: AppendixB_1L}
In order to obtain a result for the two-loop amplitudes up to $\mathcal{O}(\epsilon^{2})$, we ought to calculate the one-loop master integrals and amplitudes up to $\mathcal{O}(\epsilon^4)$. Denoting
\begin{align}
I_{a_{1}, ..., a_{4}} = \int \frac{e^{ \epsilon \gamma_E}}{i\pi^{\frac{D}{2}}}  \frac{1}{(k)^{a_i} (k-p_1)^{a_i}(k-p_1-p_2)^{a_i}(k-p_1-p_2-p_3)^{a_i}} \, d^Dk\,,
\end{align}
a canonical basis for this process is (for simplicity we set $M^2_{H} = -1$ again)
\begin{align}
\vec{I} = \begin{pmatrix}
   - (1-y-z)\epsilon\,  I_{2,0,1,0}\\
-z\epsilon \,   I_{0,2,0,1}\\
\epsilon \, I_{2,0,0,1}\\
(1-y-z)z\epsilon^2 I_{1,1,1,1}
\end{pmatrix}\,,
\end{align}
representing 3 bubble diagrams with a squared propagator in the variables $s_{12}$, $s_{23}$ and $M_{H}^2$, respectively, and ultimately the box diagram. The former three can be related to simple bubbles by IBPs:
\begin{alignat}{2}\label{eqn: canon_bubble1}
    I_{2,0,1,0} &= -\frac{1-2\epsilon}{-1+y+z}\mathrm{Bub}_{s_{12}} &&= \frac{(1-y-z)^{-1-\epsilon}}{\epsilon}C(\epsilon)\,,\\\label{eqn: canon_bubble2}
    I_{0,2,0,1} &= \frac{1-2\epsilon}{z}\mathrm{Bub}_{s_{23}} &&= \frac{(z)^{-1-\epsilon}}{\epsilon}C(\epsilon)\,,\\\label{eqn: canon_bubble3}
    I_{2,0,0,1} &= (1-2\epsilon)\mathrm{Bub}_{m_{H}^2} &&= \frac{1}{\epsilon}C(\epsilon)\,,
\end{alignat}
where we used the well-known expression for the bubble,
\begin{align}
    \mathrm{Bub}_{s} = s^{-\epsilon}\frac{1}{\epsilon(1-2\epsilon)}\left(-e^{ \epsilon \gamma_E} \frac{\Gamma(1+\epsilon)\Gamma(1-\epsilon)^2}{\Gamma(1-2\epsilon)}\right) = s^{-\epsilon}\frac{1}{\epsilon(1-2\epsilon)}C(\epsilon)\,.
\end{align}
The differential equations take the form $\partial_y \vec{I} = A \vec{I}$ and $\partial_z \vec{I} = B \vec{I}$ where

  \begin{align}
    A&= \begin{pmatrix}
\frac{1}{x} & 0 & 0 & 0\\
0 & 0 & 0 & 0\\
0 & 0 & 0 & 0\\
\left(\frac{2}{y} - \frac{2}{y+z}\right) & \frac{2}{y} &\left(-\frac{2}{y} + \frac{2}{y+z}\right) &\left( \frac{1}{y} + \frac{1}{x}\right)
\end{pmatrix}\;,\\[1em]
    B&= \begin{pmatrix}
\frac{1}{x} & 0 & 0 & 0\\
0 & -\frac{1}{z} & 0 & 0\\
0 & 0 & 0 & 0\\
 -\frac{2}{y+z} & \frac{-2}{-1+z} &\left(\frac{2}{-1+z} + \frac{2}{y+z}\right) &\left(-\frac{1}{z} + \frac{1}{x}\right)
\end{pmatrix}
  \end{align}

and $x = 1-y-z$. They confirm the scaling of the bubbles given in (\ref{eqn: canon_bubble1}, \ref{eqn: canon_bubble2}, \ref{eqn: canon_bubble3}), and allow us to compute the box. The boundary condition obtained by multiplying the first equation by $y$ and sending it to 0 reads
\begin{align}\label{eqn: BC}
    0 = 2 I_{2,0,1,0}\big|_{y=0} + 2 I_{0,2,0,1}\big|_{y=0}  -2 I_{2,0,0,1}\big|_{y=0} + I_{1,1,1,1}\big|_{y=0}
\end{align}
and is the only condition needed to find a unique solution for the box integral. We could have obtained this requirement also via considering the limit $y\to0$ for $\epsilon<0$. The solution to the DE in $y$ near the point $y=0$ is 
\begin{align}
e^{\epsilon \log{y} \lim_{y\to0}[yA_{y}]} \vec{I}\big|_{y=0} = 
\begin{pmatrix}
    1 & 0 & 0 & 0\\
0 & 1 & 0 & 0\\
0 & 0 & 1 & 0\\
 2(-1+y^\epsilon) &  2(-1+y^\epsilon) & -2(-1+y^\epsilon) &  y^\epsilon
\end{pmatrix}\vec{I}\big|_{y=0}\,.
\end{align}
Clearly requiring the non-appearance of terms $y^\epsilon$ in the solution replicates the condition \eqref{eqn: BC}. The strength of the second approach is that it applies to regular as well as singular limits. Expanded solutions for all four integrals can be found in the supplementary material.

\bibliographystyle{JHEP}
\bibliography{main}
\end{document}